\documentclass[aps,prd,twocolumn,showpacs,nofootinbib,amsmath,amssymb,floatfix,superscriptaddress,showkeys]{revtex4}
\usepackage{graphicx}
\usepackage{dcolumn}
\usepackage{bm}
\usepackage{captcont}
\usepackage{xcolor}

\usepackage{epstopdf}
\usepackage{mathtools}
\usepackage{natbib}
\usepackage{ulem}
\definecolor{blue0}{rgb}{0,0,0.6}
\usepackage[colorlinks,linkcolor=blue0,anchorcolor=blue0,citecolor=blue0,urlcolor=blue0]{hyperref}

\newcommand{\jcap}{J. Cosmol. Astropart. Phys.}

\newcommand{\aap}{Astron. Astrophys}

\begin{document}
\title{Searching for spectral oscillations due to photon-axionlike particle conversion using the Fermi-LAT observations of bright supernova remnants}

\author{Zi-Qing Xia}
\affiliation{Key Laboratory of Dark Matter and Space Astronomy, Purple Mountain Observatory, Chinese Academy of Sciences, Nanjing 210008, China}
\affiliation{School of Astronomy and Space Science, University of Science and Technology of China, Hefei, Anhui 230026, China}
\author{Cun Zhang}
\affiliation{Key Laboratory of Dark Matter and Space Astronomy, Purple Mountain Observatory, Chinese Academy of Sciences, Nanjing 210008, China}
\affiliation{School of Physics, Nanjing University, Nanjing, 210092, China}
\author{Yun-Feng Liang}
\email{Corresponding author. liangyf@pmo.ac.cn}
\affiliation{Key Laboratory of Dark Matter and Space Astronomy, Purple Mountain Observatory, Chinese Academy of Sciences, Nanjing 210008, China}
\author{Lei Feng}
\email{Corresponding author. fenglei@pmo.ac.cn}
\affiliation{Key Laboratory of Dark Matter and Space Astronomy, Purple Mountain Observatory, Chinese Academy of Sciences, Nanjing 210008, China}
\author{Qiang Yuan}
\email{Corresponding author. yuanq@pmo.ac.cn}
\affiliation{Key Laboratory of Dark Matter and Space Astronomy, Purple Mountain Observatory, Chinese Academy of Sciences, Nanjing 210008, China}
\affiliation{School of Astronomy and Space Science, University of Science and Technology of China, Hefei, Anhui 230026, China}
\author{Yi-Zhong Fan}
\email{Corresponding author. yzfan@pmo.ac.cn}
\affiliation{Key Laboratory of Dark Matter and Space Astronomy, Purple Mountain Observatory, Chinese Academy of Sciences, Nanjing 210008, China}
\affiliation{School of Astronomy and Space Science, University of Science and Technology of China, Hefei, Anhui 230026, China}
\author{Jian Wu}
\affiliation{Key Laboratory of Dark Matter and Space Astronomy, Purple Mountain Observatory, Chinese Academy of Sciences, Nanjing 210008, China}
\affiliation{School of Astronomy and Space Science, University of Science and Technology of China, Hefei, Anhui 230026, China}

\date{\today}

\begin{abstract}
Axionlike-particles (ALPs) are one promising type of dark matter candidate particle that may generate detectable effects on $\gamma$-ray spectra other than the canonical weakly interacting massive particles. In this work we search for such oscillation effects in the spectra of supernova remnants caused by the photon-ALP conversion, using the Fermi Large Area Telescope data. Three bright supernova remnants, IC443, W44, and W51C, are analyzed. 
The inclusion of photon-ALP oscillations yields an improved fit to the $\gamma$-ray spectrum of IC443, which gives a statistical significance of $4.2\sigma$ in favor of such spectral oscillation. 
However, the best-fit parameters of ALPs ($m_{a}=6.6\,{\rm neV}$, $g_{a\gamma}=13.4 \times 10^{-11}\,{\rm GeV}^{-1}$) are in tension with the upper bound ($g_{a\gamma}< 6.6 \times 10^{-11}\,{\rm GeV}^{-1}$) set by the CAST experiment.
It is difficult to explain the results using the systematic uncertainties of the flux measurements.
We speculate that the ``irregularity" displayed in the spectrum of IC443 may be due to the superposition of the emission from different parts of the remnant.
\end{abstract}
\pacs{95.35.+d, 95.85.Pw, 98.58.Mj}
\keywords{Dark matter$-$Gamma rays: general$-$ISM: supernova remnants}

\maketitle

\section{Introduction}
\label{sec1}

The presence of a large amount of dark matter (DM) in the Universe has already been convincingly established. Due to the very weak interaction of DM particles with standard model particles, the particle nature of DM is still far from clear. At present, the most popular DM candidate is a kind of weakly interacting massive particle, which can naturally account for the DM density in the current Universe assuming that they decoupled from a plasma soup in the very early Universe. 
Axion, a hypothetical sub-eV particle beyond the standard model introduced to solve the $CP$ violation in strong interaction \cite{pq1977,Weinberg1978,axion78,axion90}, is one promising candidate of cold DM. Because the axion mass and the coupling to photons are correlated with each other, the conventional scenario of axions has been strongly constrained or ruled out by ground-based experiments \cite{CAST,alpex,alpex1}. Axionlike-particles (ALPs), particles that have similar properties to axions, have been proposed as an alternative to axions to explain the DM. The mass and coupling of ALPs can be independent from each other, which can avoid such experimental constrains. ALPs may make up a significant fraction or all of the DM in the Universe \cite{ALPdm}.

The Lagrangian of the interaction between photons and ALPs in the magnetic field can be written as
$$\mathcal{L}=g_{a\gamma}\vec{E}\cdot\vec{B}{a},$$
where $\vec{E}$, $\vec{B}$, and $a$ are electric, magnetic, and axionlike fields, and $g_{a\gamma}$ is the coupling constant.
High-energy $\gamma$-ray photons can oscillate to ALPs (and vice versa) when passing through external magnetic fields, if the field strength is strong enough and the propagation distance is long enough. The photon-ALP oscillation induces specific modulation in the spectra of $\gamma$-ray sources, which could be a smoking-gun signature for the existence of ALPs \cite{Hochmuth07,Hooper07,Angelis08,Simet08,sc09,Meyer13,Ayala14}.

The $\gamma$-ray data have been widely used to search for ALP signals \cite{belikov11alp,hess13pks2155,reesman14alp,berenji16alp_ns,fermi16alp,meyer17sne,zc16alp,Majumdar17psr1,ALP2018,Majumdar17psr2}. The Fermi Collaboration searched for ALPs from the radio galaxy NGC 1275, located in the Persus cluster, which has relatively good magnetic field measurements. No significant ALP signal was found, and a large part of the parameter space for the low $\gamma$-ray opacity of the Universe was excluded \cite{fermi16alp}. 
The HESS Collaboration used the data of PKS 2155-304 to constrain the ALP properties, and derived an upper limit of the ALP-photon coupling at the 95\% confidence limit (CL) to be $g_{\gamma{a}}<2.1\times10^{-11}\,{\rm GeV^{-1}}$ for ALP masses of $15-60\,{\rm neV}$. With the Fermi Large Area Telescope (Fermi-LAT) observation of PKS 2155-304, Zhang $et al.$ \cite{zc16alp} showed that the holelike feature that survived in the constraints of NGC 1275 by the Fermi Collaboration \cite{fermi16alp} can be further excluded.
Analysis of Galactic sources has also been used to search for ALPs. By studying modulation behaviors in spectra of 6 $\gamma$-ray pulsars, Majumdar $et al.$ \cite{Majumdar17psr1,ALP2018} reported intriguing indications of photon-ALP mixing. Even stronger evidence was obtained in a further analysis of 12 non-Galactic plane pulsars \cite{Majumdar17psr2}. 
In addition, a recent research used the photons emitted by high-energy neutrino sources to search for axionlike particles and found that an experiment like LHAASO could probe very deep into the ALP parameter space \cite{ALP2017LHAASO}.

In this work we search for possible ALP-photon oscillation signals in $\gamma$-ray spectra of supernova remnants (SNRs). Note that $\gamma$-ray emission of SNRs is from either the inverse Compton radiation of high-energy electrons accelerated in SNR shocks or the inelastic collision between high-energy protons and the surrounding medium. Within all the sources observed by the Fermi-LAT, SNRs are among the ones with highest fluxes \cite{3fgl, fermi16snrcata} and are thus good targets for the search for spectral distortions. Moreover, SNRs are usually located in the Galactic plane, and the magnetic fields along the lines of sight are relatively high. 

We focus on three middle-aged SNRs, IC 443, W44 and W51C. They are the brightest SNRs observed by the Fermi-LAT \cite{fermi16snrcata}. Evidence of $\pi^0$-decay emission is displayed in their $\gamma$-ray spectra \cite{fermi13pi0,Jogler16w51c}. The basic information of these sources, including their positions in the sky and distances, is listed in Table. \ref{tb1}.

\section{Photon-ALP Oscillation in The Milky Way Magnetic Field}
\label{sec2}

In the Milky Way magnetic field, photons and ALPs can oscillate into each other. For a homogeneous magnetic field with size $l$, the survival probability for an initially polarized photon with energy $E_\gamma$ is \cite{axionf,axionf1}
\begin{eqnarray}
P_{\rm ALP}&=&1-P_{\gamma \rightarrow a} \\ \nonumber
&=&1-\frac{1}{1+{E_{\rm c}^2}/{E_\gamma^2}} \sin^2 \left[\frac{g_{a\gamma} B_{\rm T} l}{2}  \sqrt{1+\frac{E_{\rm c}^2}{E_\gamma^2}} \right],
\end{eqnarray}
where the characteristic energy $E_{\rm c}$ is defined as
\begin{eqnarray}
E_{\rm c}=\frac{\left| m_a^2-w_{\rm pl}^2\right|}{2g_{a\gamma} B_{\rm T}},
\end{eqnarray}
and $w_{\rm pl}^2=4\pi\alpha n_e/m_e$ is the plasma frequency.
Strong mixing between photons and ALPs happens when the photon energy is higher than the characteristic energy. With $B_T \sim 1\,\mu {\rm G}$, $n_e=10^{-1} {\rm cm^{-3}}$, $g_{a\gamma}=10^{-10}\,{\rm GeV}^{-1}$, and $m_a=10^{-9}\,{\rm eV}$, the characteristic energy is about $250~{\rm MeV}$, which is in the energy range of Fermi-LAT.
To calculate the survival probability in the Milky Way magnetic field, we solve the evolution equation for the photon-ALP beam as in Refs. \cite{axionf,axionf1}. For simplicity, the $\gamma$-rays from the source are assumed to be unpolarized. Such an assumption would yield conservative results.

\begin{figure}
\centering
\includegraphics[width=0.45\textwidth]{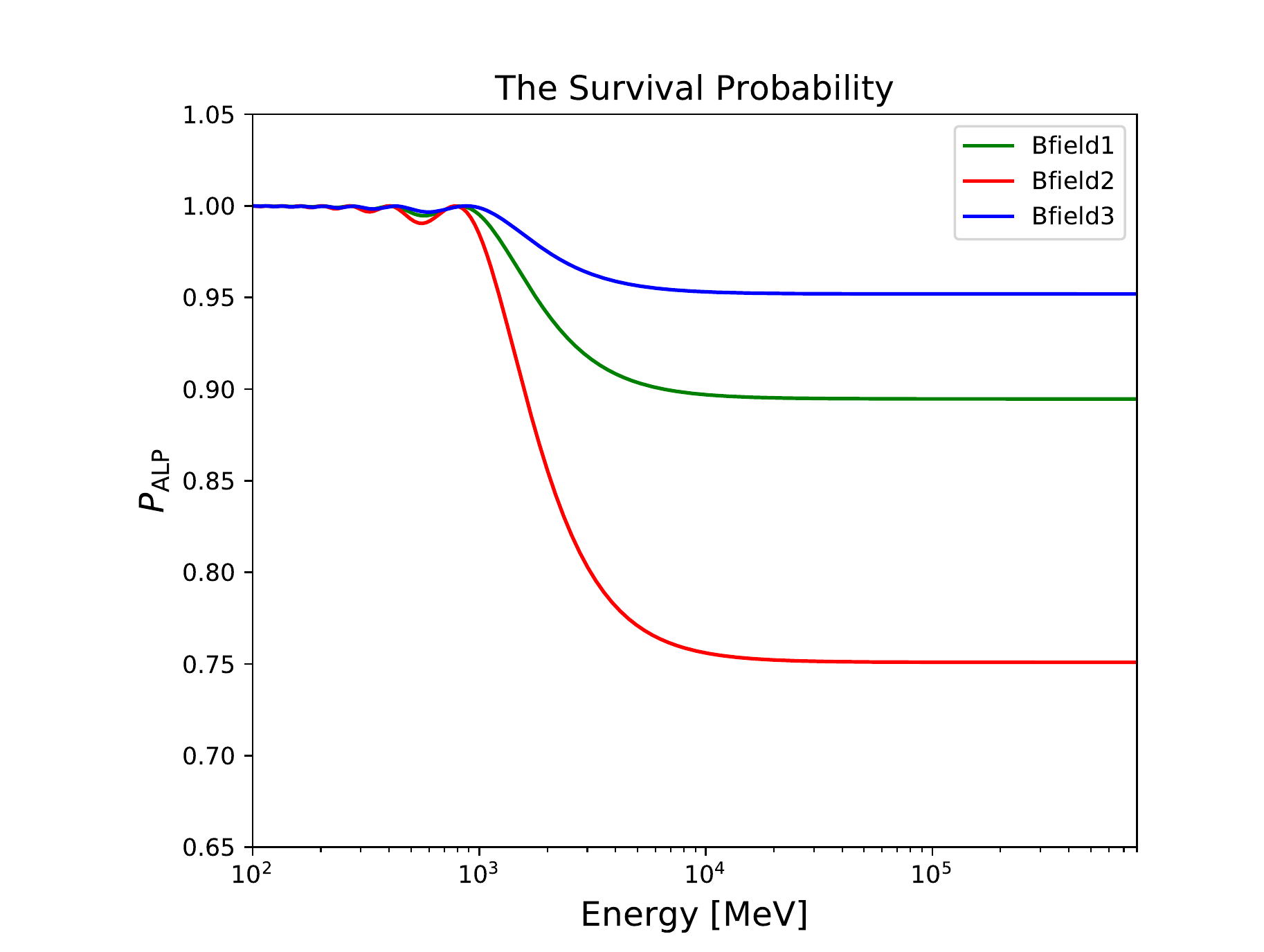}
\caption{Survival probabilities $P_{\rm ALP}$ for photons of IC443, for 
$(m_{a},g_{a\gamma})=(6.6~{\rm neV},13.4 \times 10^{-11}~{\rm GeV}^{-1})$, and three magnetic field models.}
\label{fig:PALP}
\end{figure}

The magnetic field of the Milky Way consists of a large-scale regular component and a small-scale random component. The coherence length of the random magnetic field is much smaller than the regular magnetic field as well as the photon-ALP oscillation length. Therefore, we neglect the random magnetic field.
For the regular magnetic field, we consider three different models developed by Jansson and Farrar \cite{Bfield1}, Sun $et al.$ \cite{Bfield2}, and Pshirkov $et al.$ \cite{Bfield3}, which are denoted as Bfield1, Bfield2 and Bfield3, respectively. Figure \ref{fig:PALP} illustrates the survival probabilities for photons of IC443, for given ALP parameters and the three magnetic field models.

\begin{table*}[t]
\caption{The basic information and ALP analysis results}
\begin{ruledtabular}
\begin{tabular}{lrrrrccccc}
  Name & {\it \rm lon [$^{\rm \circ}$]} & {\it \rm lat [$^{\rm \circ}$]} & {\rm d [${\rm kpc}$]} & {\it $ \rm  TS\footnote{TS = $ {\chi^2_{\rm w/oALP,min} - \chi^2_{\rm wALP,min}}$}_{Bfield1}$} &{\it $ \rm TS_{Bfield2}$} &{\it  $ \rm TS_{Bfield3} $}\\
\hline
IC443 &189.065 & 3.235 & 1.5  & 21.2 & 21.5 & 21.8 \\[3pt]
W44 & 34.560 & -0.497 & 3  & 4.6 & 3.9 & 3.7 \\[3pt]
W51C & 49.131 & -0.467 & 5.5  &  2.2 & 3.1 & 5.0 \\[3pt]
\hline
Geminga & 195.133 & 4.270 & 0.25 & 9.1 & 9.2 & 6.3 \\[3pt] 
Vela & 263.555 & -2.787 & 0.3 & 0.9 & 0.3 & 1.9 \\[3pt] 
\end{tabular}
\end{ruledtabular}
\label{tb1}
\end{table*}

\section{LAT DATA ANALYSES}
\label{sec3}

We use nearly nine years of Fermi-LAT {\tt Pass 8} data from October 27, 2008 (MET = 246823875) to August 15, 2017 (MET = 524448005). The {\tt Pass 8} data have several important improvements compared with previous versions, including an extension of the energy range, better energy measurements, and a larger effective area \cite{pass8econf}. We select the photons with energies from 300 MeV to 800 GeV. The {\tt FRONT+BACK} conversion-type data with the {\tt SOURCE} event class are adopted in the analysis. We apply a zenith angle cut of ${\rm \theta} < 90^{\rm \circ}$ to reduce the contribution from the Earth's limb and adopt the recommended quality-filter cuts {\tt(DATA\_QUAL==1 \&\& LAT\_CONFIG==1)} to extract the good time intervals. The instrument response functions (IRFs) {\tt P8R2\_SOURCE\_V6} and the diffuse emission templates {\tt gll\_iem\_v06.fits} and {\tt iso\_P8R2\_SOURCE\_V6.txt} are used, which are available from the Fermi Science Support Center\footnote{\url {http://fermi.gsfc.nasa.gov/ssc/}}.

The standard binned likelihood analysis is performed for 30 evenly spaced logarithmic energy bins. When the TS (test statistic, defined as two times the logarithmic likelihood ratio between the signal hypothesis and the null hypothesis of the target source) value of the target SNR is smaller than 25, the {\tt pyLikelihood UpperLimits} tool is adopted to calculate the 95\% upper limit of the flux in that energy bin. 

Then we perform the $\chi^2$ analysis on the obtained spectral energy distribution (SED). The intrinsic spectrum of the SNR is modeled by a {\tt LogParabola} function\footnote{\url {https://fermi.gsfc.nasa.gov/ssc/data/analysis/scitools/source\_models.html}}, as used in the Third Fermi-LAT catalog (3FGL) \cite{3fgl}
\begin{equation}
{{\left(\frac{dN}{dE}\right)}_{\rm SNR}=N_0\left(\frac{E}{E_{\rm b}}\right)^{-[\alpha+\beta {\rm log}(E/E_{\rm b})]} },
\label{eq:LogParabola}
\end{equation}
where $E_{\rm b}$ is a scale parameter which is fixed \cite{logp}\footnote{We have tested that setting $E_{\rm b}$ free only affects the results slightly.}.

To take into account the photon-ALP oscillation effect, we multiply the intrinsic spectrum by the survival probability $P_{\rm ALP}$. The energy spectrum with photon-ALP oscillation can be expressed as
\begin{equation}
{{\left(\frac{dN}{dE}\right)}_{\rm wALP}=P_{\rm ALP}(g_{a\gamma}, m_{a}, E){\left(\frac{dN}{dE}\right)}_{\rm SNR}}.
\label{eq:ALP}
\end{equation}
We further consider the energy dispersion of Fermi-LAT. After convolving the energy dispersion function $D_{\rm eff}(E',E)$, we get the final observed spectrum
\begin{equation}
{\frac{dN}{dE'}}=D_{\rm eff}\otimes{\frac{dN}{dE}},
\label{eq:Deff}
\end{equation}
where $E'$ and $E$ represent the observed and the true energy.

The models with or without the photon-ALP oscillation are used to fit the Fermi-LAT spectra. The fits are repeated for a grid of ALP masses $m_{a}$ and photon-ALP couplings $g_{a\gamma}$. The ALP parameters are fixed in each fit. Meanwhile, for both models, $N_0$, $\alpha$, and $\beta$ are set to be free, while $E_{\rm b}$ is frozen to the default value in the 3FGL \cite{3fgl}. By scanning a grid of ($m_{a}$, $g_{a\gamma}$), we obtain the optimal ALP parameters in the parameter space. The scan ranges of the coupling constant and the ALP mass are $0.1-100 \times 10^{-11}\,{\rm GeV}^{-1}$ and $0.1-100\,{\rm neV}$, respectively.

\section{RESULTS}
\label{sec4}

\begin{figure*}
\centering
\includegraphics[width=0.45\textwidth]{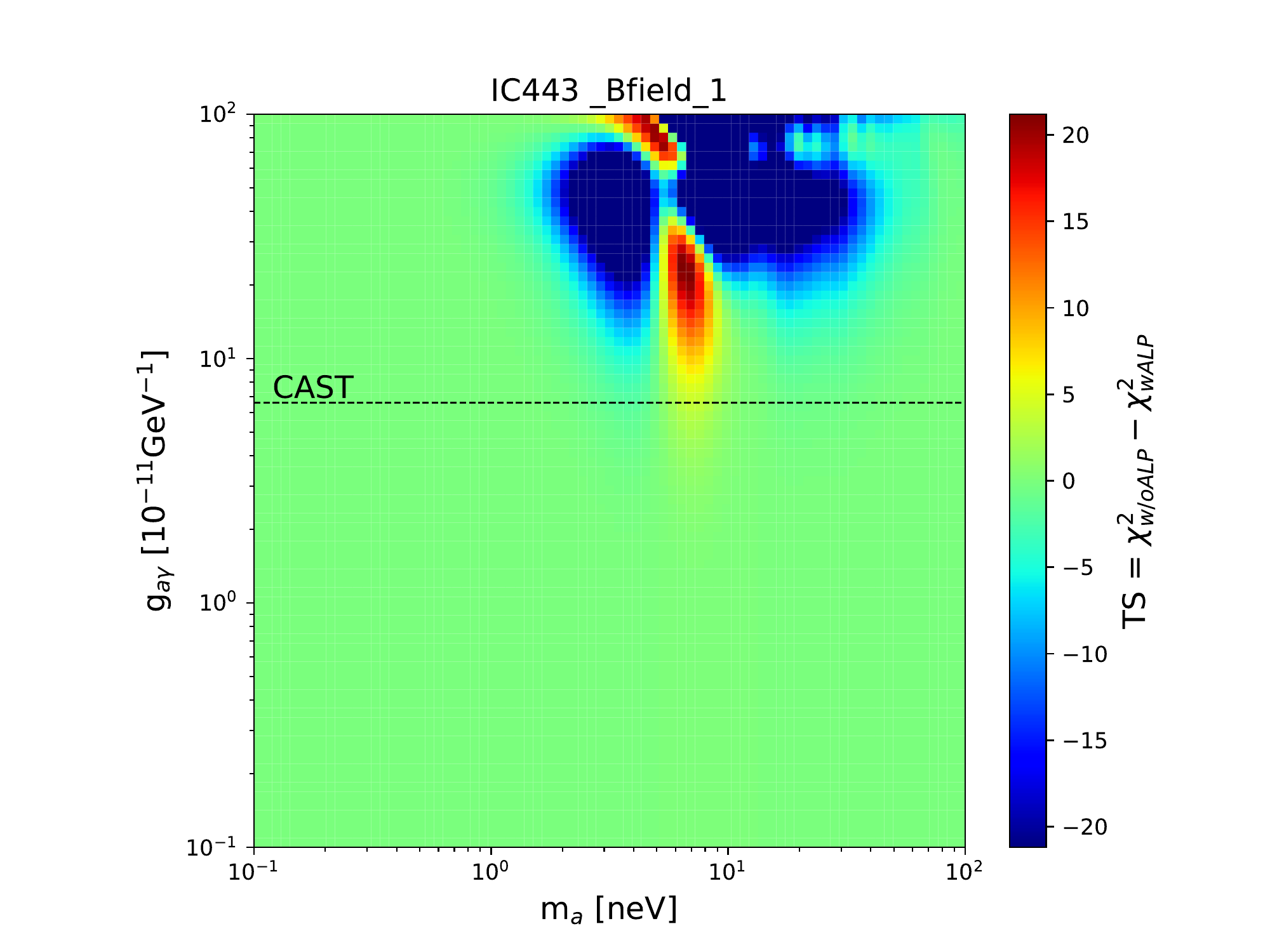}
\includegraphics[width=0.45\textwidth]{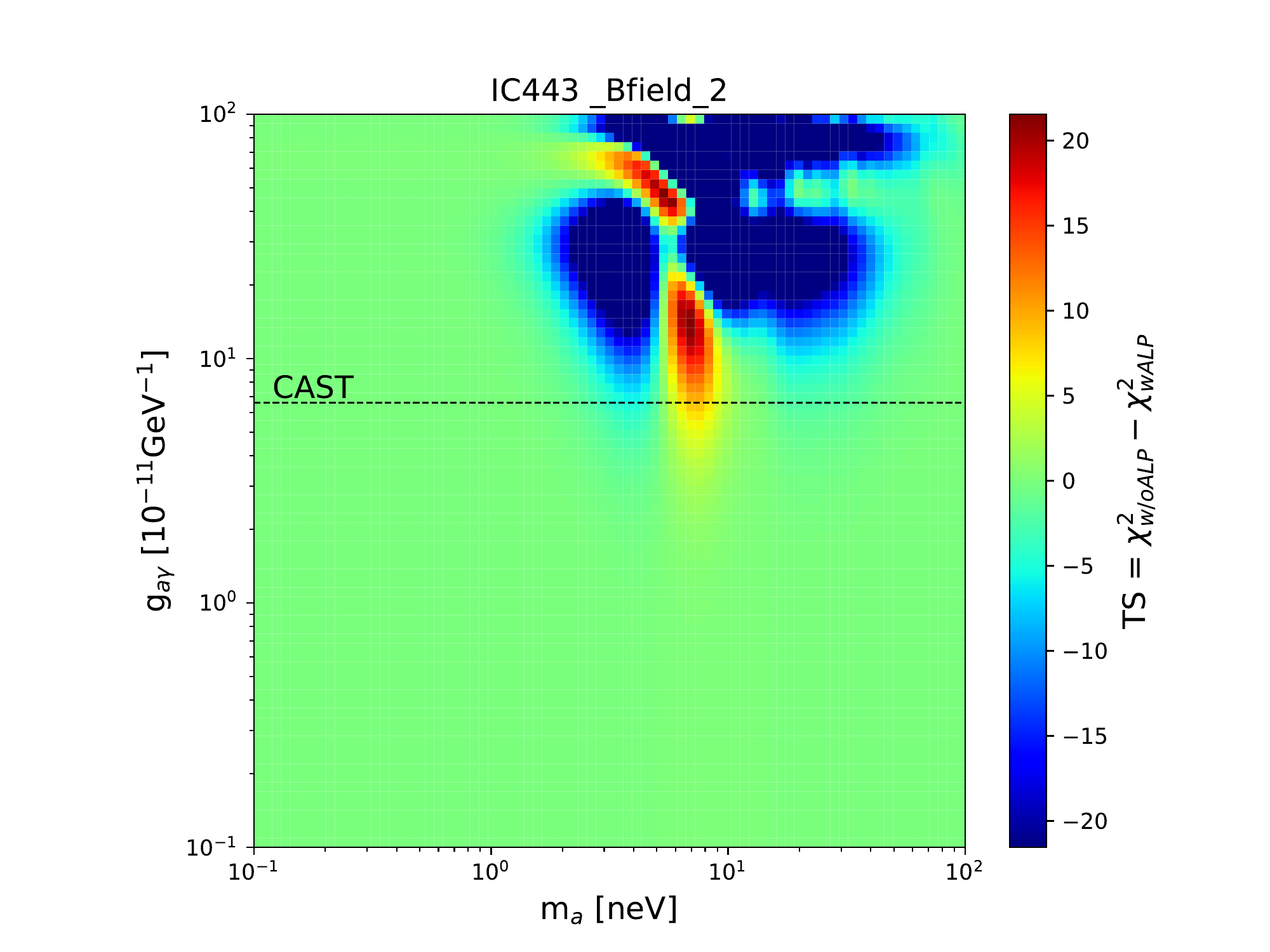}
\includegraphics[width=0.45\textwidth]{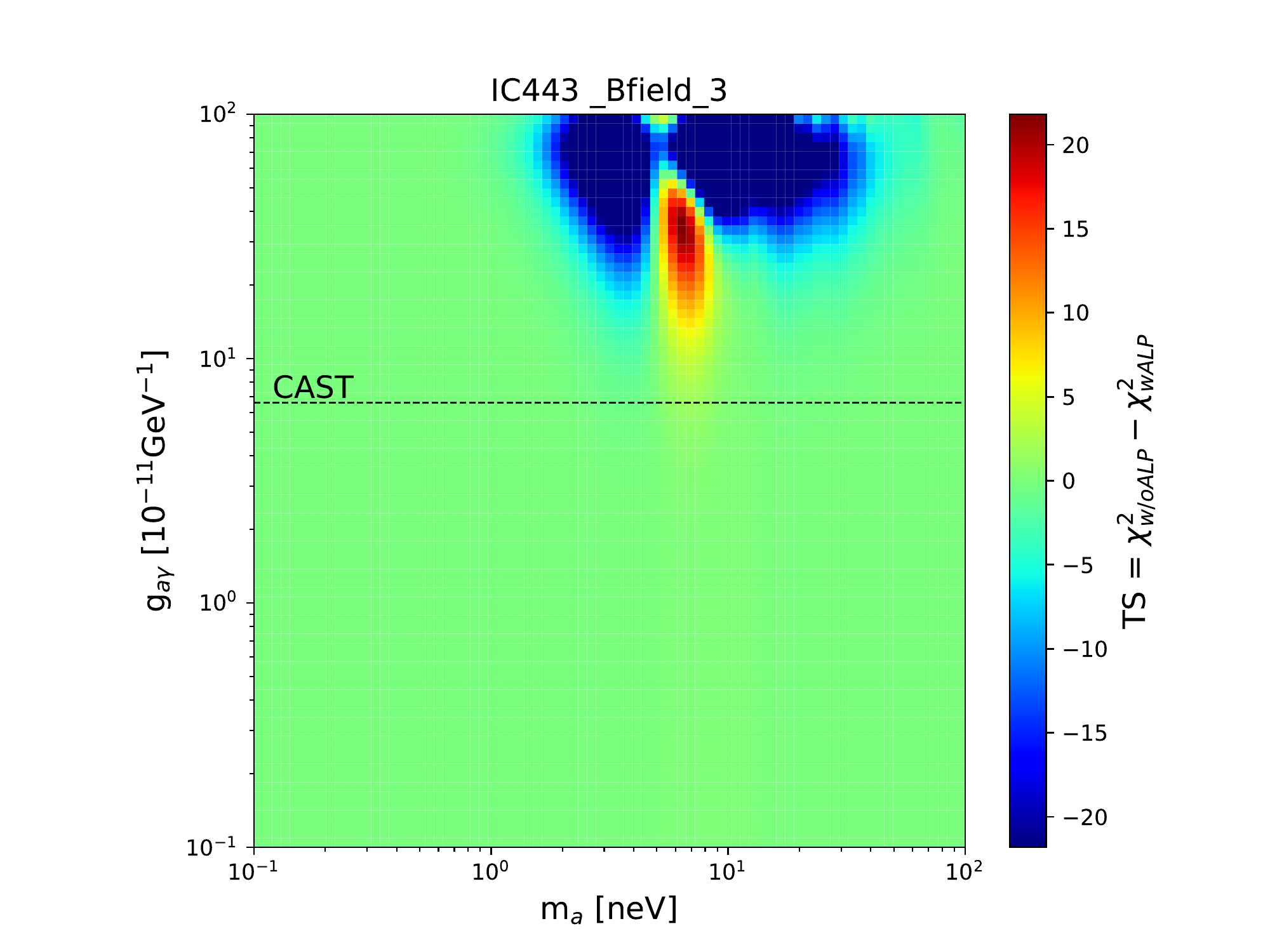}
\caption{The TS value as a function of ALP mass $m_{a}$ and photon-ALP coupling constant $g_{a\gamma}$ for IC443. Three subpanels are for three different magnetic field models. 
For better visualization we set the upper boundary of the color bar scale to the TS$_{\rm max}$ in the scanning and truncate the lower one to $-$TS$_{\rm max}$. 
The black dashed line shows the upper limit of the photon-ALP coupling set by CAST \cite{CAST}.}
\label{fig:IC443bin30}
\end{figure*}

\begin{figure}
\centering
\includegraphics[width=0.45\textwidth]{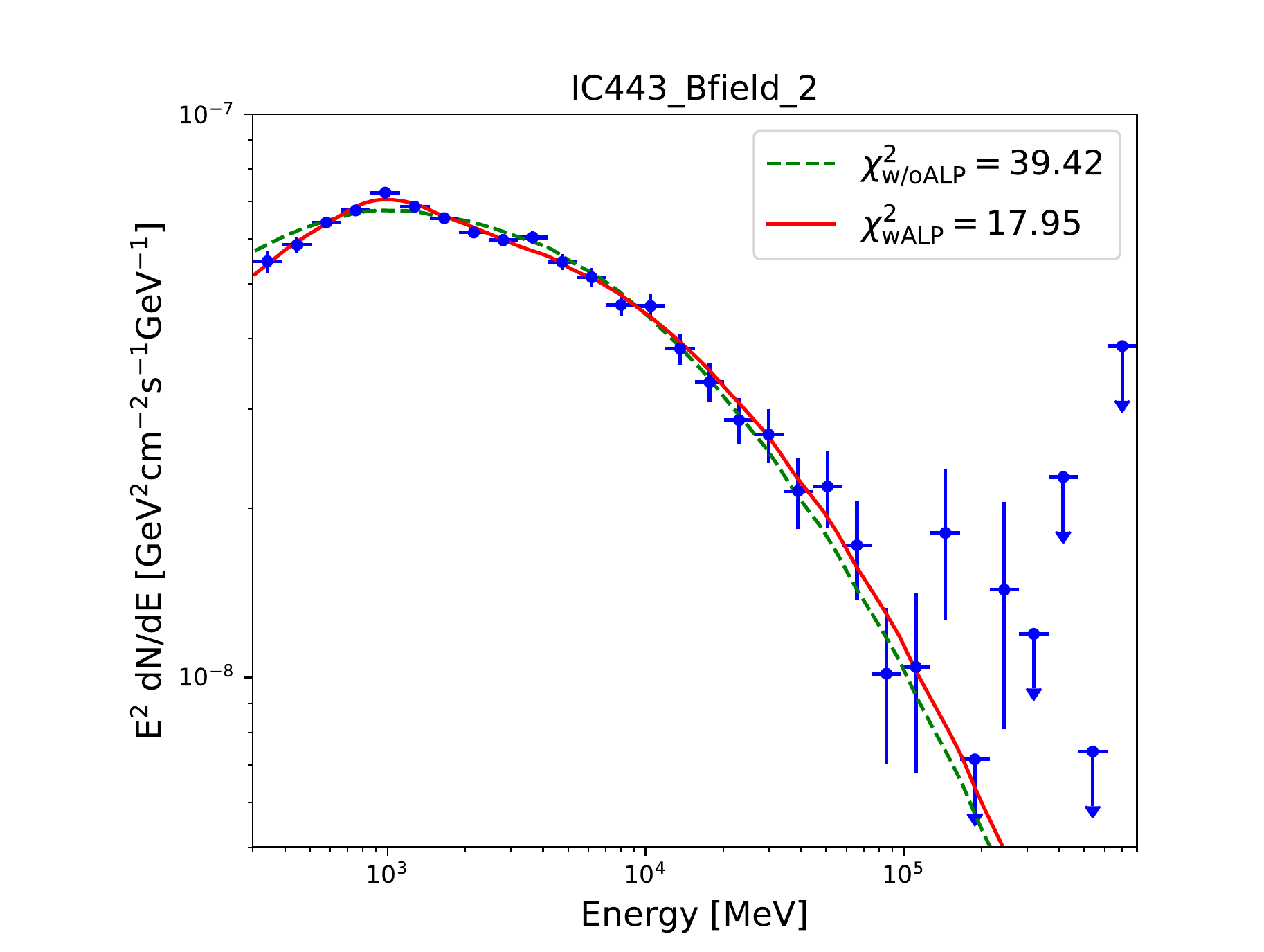}
\caption{The Fermi-LAT observed SED of IC443, compared with the best-fit models without (dashed) and with (solid) photon-ALP oscillation, for Bfield2.}
\label{fig:IC443SED}
\end{figure}

The fitting TS values of the ALP model (TS = ${\chi^2_{\rm w/oALP} - \chi^2_{\rm wALP}}$) for the three SNRs and three magnetic field models are tabulated in Table \ref{tb1}. 
We find that IC443 gives a relatively high TS value ($\sim21$) of the oscillation model. The TS values depend weakly on the chosen magnetic field model, indicating that the presence of spectral irregularities is robust. For the other two SNRs, the preference of photon-ALP oscillation is not significant.

The TS values of the two-dimensional parameter space ($m_{a}$, $g_{a\gamma}$) for IC443 are shown in Fig. \ref{fig:IC443bin30}. The red (blue) regions are favored (disfavored) by the data. 
In the case of the Bfield2, we get an improvement of about 21.5 for the $\chi^2$ (i.e., ${\chi^2_{\rm w/oALP} - \chi^2_{\rm wALP}}=21.5$). With 2 degrees of freedom, this means a local significance of $4.2\sigma$. The best-fit values of the coupling constant and ALP mass are $g_{a\gamma}=13.4 \times 10^{-11}~{\rm GeV}^{-1}$ and $m_{a}=6.6~{\rm neV}$, respectively. The comparison between the models and the observational SED is given in Fig. \ref{fig:IC443SED}. We can see that including the photon-ALP oscillation, the model prediction matches the data better.

Nevertheless, the CAST experiment set the upper limit of the coupling constant $g_{a\gamma}$ as $ 6.6  \times 10^{-11}\,{\rm GeV}^{-1}$ (the black dashed line in Fig. \ref{fig:IC443bin30}) \cite{CAST}. The best-fit values of the coupling constant for IC443 for the three kinds of Galactic magnetic field models are all excluded by the CAST limits. This challenges the photon-ALP oscillation interpretation of the spectral irregularities of the $\gamma$-ray spectrum of IC443.

\section{SYSTEMATIC UNCERTAINTIES}
\label{sec5}

\subsection{Binning of the data}
\begin{figure}
\centering
\includegraphics[width=0.45\textwidth]{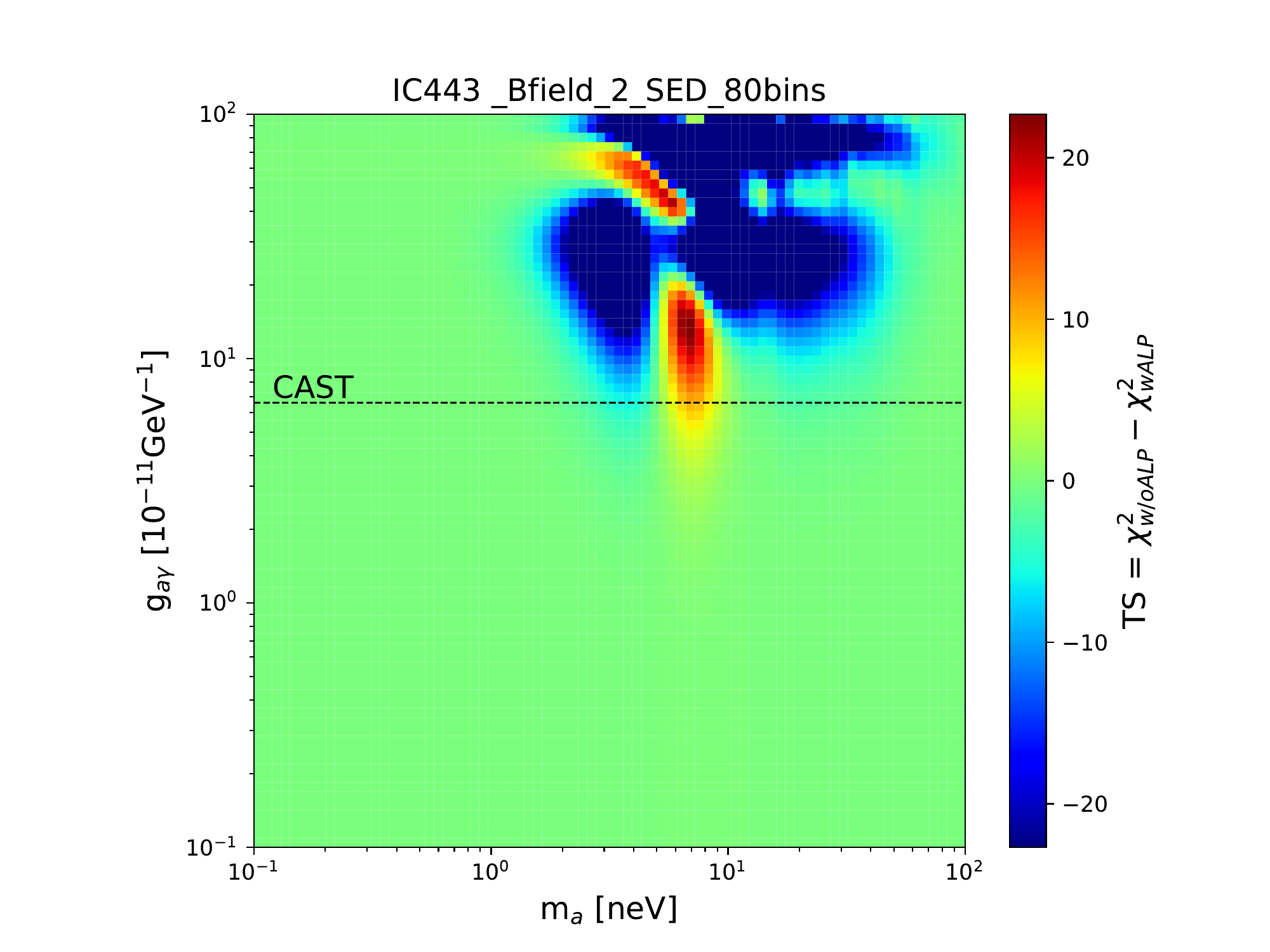}
\caption{The TS value as a function of ALP mass $m_{a}$ photon-ALP coupling constant $g_{a\gamma}$ for IC443 with 80 energy bins, for the case of Bfield2.}
\label{fig:IC443bin80}
\end{figure}

To test the possible binning effect, we repeat the analysis of IC443 for 80 evenly spaced logarithmic energy bins from 300 MeV and 800 GeV. The TS distribution as a function of $m_{a}$ and $g_{a\gamma}$ is plotted in Fig. \ref{fig:IC443bin80}. We find that the result is quite consistent with the previous one. The TS value of the best-fit model becomes slightly larger than that of 30 bins (for example, the TS value is 22.7 for Bfield2, compared with 21.5 for 30 bins), likely due to the fact that the oscillation effect is narrower than the widths of  the energy bins.

\subsection{Instrument performance}

In order to estimate the systematic uncertainties from the instrument performance \cite{Majumdar17psr1,ALP2018,Majumdar17psr2}, we repeat the analysis for two of the brightest pulsars, Geminga and Vela. Both pulsars are very close, and the propagation distances of $\gamma$-ray photons are so small that the expected photon-ALP oscillation effects are weak. 
Furthermore, Geminga is located just $6^{\rm \circ}$ away from IC443, and the instrumental performance for Geminga should be similar to that for IC443. Therefore, Geminga is a good object to evaluate the systematic uncertainties on the observations due to instrument performance. The result of Vela is adopted as a further cross-check.

\begin{figure*}[t]
\centering
\includegraphics[width=0.45\textwidth]{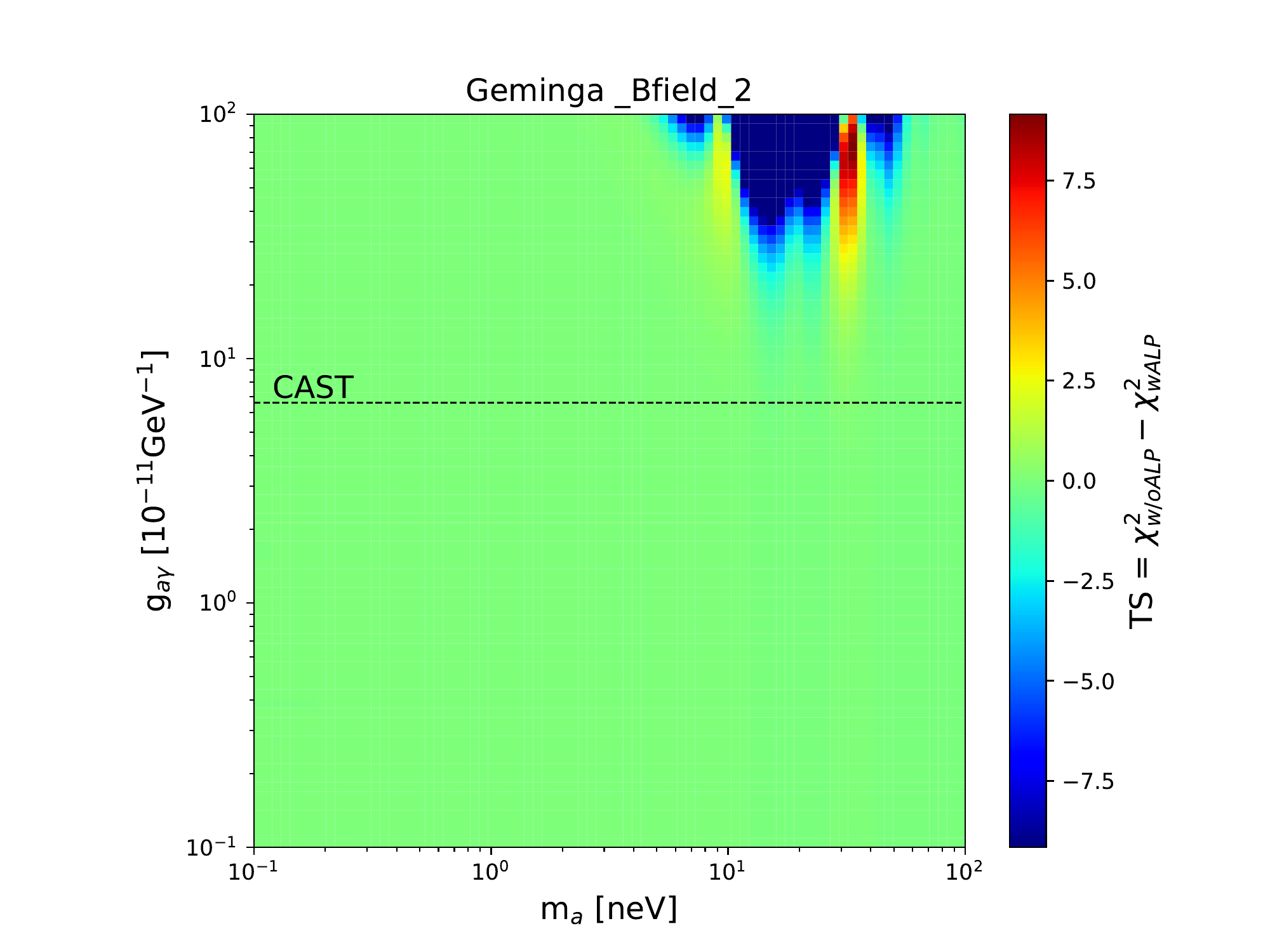}
\includegraphics[width=0.45\textwidth]{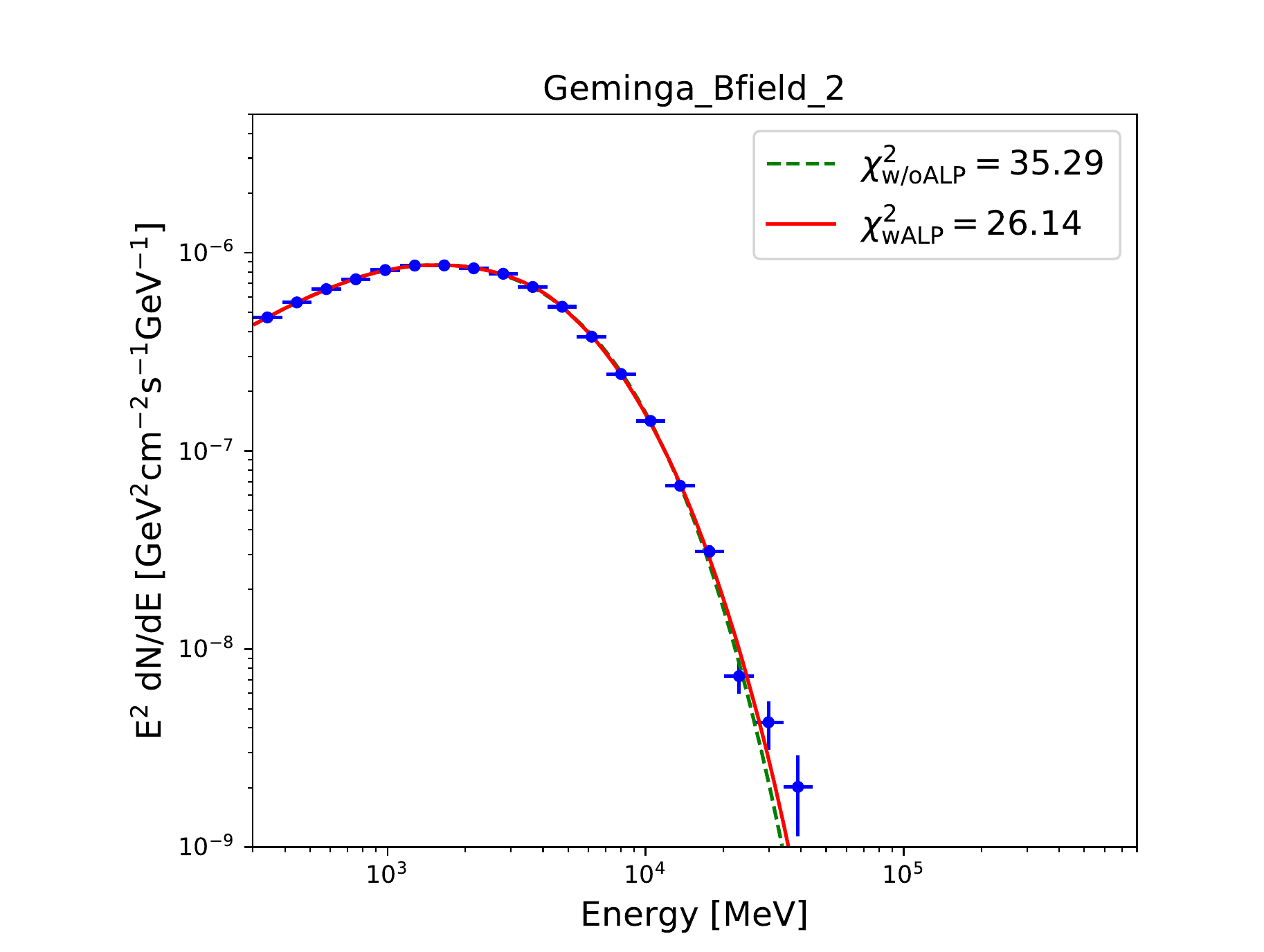}
\caption{Left panel: The TS value as a function of ALP mass $m_{a}$ and photon-ALP coupling constant $g_{a\gamma}$ for Geminga. Right panel: The SED of Geminga and the best-fit spectra without and with photon-ALP oscillations. Bfield2 is adopted.}
\label{fig:Geminga}
\end{figure*}

The intrinsic spectrum of Geminga is modeled by a power law with a superexponential cutoff {\tt PLSuperExpCutoff}\footnote{\url {https://fermi.gsfc.nasa.gov/ssc/data/analysis/scitools/source\_models.html}}
\begin{equation}
{\left(\frac{dN}{dE}\right)_{\rm pulsar}=N_0\left(\frac{E}{E_{\rm 0}}\right)^{\gamma_{\rm1}}{\rm exp}\left[-\left(\frac{E}{E_{\rm c}}\right)^{\gamma_{\rm 2}}\right]}.
\label{eq:PLSuperExpCutoff}
\end{equation}
The scale parameter ${E_{\rm 0}}$ is again fixed to be the value from 3FGL \cite{3fgl}, while other parameters ($N_0$, $ {E_{\rm c}}$, ${\gamma_{\rm1}}$, ${\gamma_{\rm 2}}$) are left free in the fit.

The TS values are found to be about $6\textendash9$ for the three magnetic field models, as shown in Table \ref{tb1}. Figure \ref{fig:Geminga} shows the TS value distribution on the $(m_a,g_{a\gamma})$ plane (left) and the best-fit results of the spectra (right) of Geminga, for Bfield2. 
The result indicates that the obtained SED of Geminga is well described by the assumed intrinsic spectral model, and no obvious photon-ALP oscillation is found.
For the case of Vela, the TS values shown in Table. \ref{tb1} and Fig. \ref{fig:Other Results} are even smaller.
Hence we believe that the spectral irregularity of IC443 should not be due to instrumental effects.

For the sake of conservation, we also introduce 3\% systematic uncertainty of the effective area into the analysis, as recommended by the Fermi-LAT Collaboration\footnote{\url {https://fermi.gsfc.nasa.gov/ssc/data/analysis/LAT\_caveats.html}}. We find that the TS value decreases to 13.4 for Bfield2, which corresponds to a statistical significance of $3.2\sigma$.

Furthermore, there are still large sources of uncertainties on the prediction of the fluxes, such as residual particle background, energy resolution, the point-spread function, and uncertainties on energy scale calibration, which have been discussed in detail in \cite{ALP2018}. In order to quantify the whole systematic uncertainty from the instrument performance, we repeat the analysis for Geminga with a relative systematic uncertainty on the flux and artificially increase it until the resulting ${\chi^2}$ per degree of freedom is $\sim1$, similar to the data-driven method in \cite{ALP2018}. The relative systematic uncertainty we get is 1.5\%, which is smaller than the 3\% systematic uncertainty of the effective area we introduced. We therefore do not carry out further corrections on our result. 

\subsection{Other systematic uncertainties}

Since IC443 is located near the Galactic plane, the model of the Galactic diffuse emission could probably affect the measurement of the spectrum. In general, the spectrum of the Galactic diffuse emission, which is smooth and continuous, barely leads to apparent oscillation of the spectrum of the target source. Furthermore, IC443 is in the direction of the Galactic anticenter. Therefore, we expect that the uncertainty of the Galactic diffuse emission would not significantly affect our result.

To examine the uncertainty from the Galactic magnetic field model, we have considered three kinds of models in the analysis. As we can see from Fig. \ref{fig:IC443bin30}, the best fit parameters depend sensitively on the magnetic field model. However, the TS values are almost unchanged. 

Finally, there are some uncertainties of the model of the intrinsic spectrum. Due to the spatial extension of IC443, the $\gamma$-ray photons of IC443 may come from several different sub-zones. The spectral irregularity may be due to the sum of radiation from different regions, which overlaps due to limited angular resolution.

\section{SUMMARY}
\label{sec6}

In this work, we have searched for possible spectral oscillations expected due to photon-ALP mixing in the Galactic magnetic field in the spectra of three bright SNRs. Assuming three representative models of the Galactic magnetic field (i.e., Bfield1, Bfield2 and Bfield3), we fit the coupling constant ($g_{a\gamma}$) and ALP mass ($m_{a}$) with the data. 
For the case of Bfield2, we find an oscillation in the spectrum of IC443 with a statistical significance of $4.2\sigma$. The best-fit ALP mass is $m_{a}=6.6\,{\rm neV}$, and the coupling is $g_{a\gamma}=13.4 \times 10^{-11}\,{\rm GeV}^{-1}$, which exceeds the upper bound ($g_{a\gamma}< 6.6 \times 10^{-11}\,{\rm GeV}^{-1}$) set by the CAST experiment \cite{CAST}. No significant spectral oscillation for the other two SNRs is found.

Some systematic uncertainties, such as the way of binning and the instrument performance are discussed in detail.
We find that the oscillation in the observed spectrum of IC443 is hardly induced by such systematic uncertainties.
It is likely that the intrinsic spectrum of IC443 itself is irregular, probably due to the sum of different radiation regions. Further efforts are required to obtain more insights in this puzzle.

Finally, we would like to point out that the Dark Matter Particle Explorer (DAMPE), a currently on-orbit space telescope for high-energy $\gamma$-ray, electron, and cosmic-ray detection with outstanding energy resolution in a wide energy range \cite{DAMPE1,DAMPE2}, may contribute significantly to the search of photon-ALP oscillations.

\begin{acknowledgments}
The data and some analysis tools used in this paper are obtained from the Fermi
Science Support Center (FSSC) provided by NASA Goddard Space Flight Center.
This work is supported by the National Key Research and Development Program 
of China (Grant No. 2016YFA0400200), the National Natural Science Foundation
of China (Grants No. 11525313, No. 11722328, No. 11773075, No. U1738210 and No. U1738136), the 100 Talents Program of
Chinese Academy of Sciences, and the Youth Innovation Promotion Association
of Chinese Academy of Sciences (Grant No. 2016288).
\end{acknowledgments}

\bibliographystyle{apsrev4-1-lyf}
\bibliography{ic443}

\begin{thebibliography}{39}%
\makeatletter
\providecommand \@ifxundefined [1]{%
 \@ifx{#1\undefined}
}%
\providecommand \@ifnum [1]{%
 \ifnum #1\expandafter \@firstoftwo
 \else \expandafter \@secondoftwo
 \fi
}%
\providecommand \@ifx [1]{%
 \ifx #1\expandafter \@firstoftwo
 \else \expandafter \@secondoftwo
 \fi
}%
\providecommand \natexlab [1]{#1}%
\providecommand \enquote  [1]{``#1''}%
\providecommand \bibnamefont  [1]{#1}%
\providecommand \bibfnamefont [1]{#1}%
\providecommand \citenamefont [1]{#1}%
\providecommand \href@noop [0]{\@secondoftwo}%
\providecommand \href [0]{\begingroup \@sanitize@url \@href}%
\providecommand \@href[1]{\@@startlink{#1}\@@href}%
\providecommand \@@href[1]{\endgroup#1\@@endlink}%
\providecommand \@sanitize@url [0]{\catcode `\\12\catcode `\$12\catcode
  `\&12\catcode `\#12\catcode `\^12\catcode `\_12\catcode `\%12\relax}%
\providecommand \@@startlink[1]{}%
\providecommand \@@endlink[0]{}%
\providecommand \url  [0]{\begingroup\@sanitize@url \@url }%
\providecommand \@url [1]{\endgroup\@href {#1}{\urlprefix }}%
\providecommand \urlprefix  [0]{URL }%
\providecommand \Eprint [0]{\href }%
\providecommand \doibase [0]{http://dx.doi.org/}%
\providecommand \selectlanguage [0]{\@gobble}%
\providecommand \bibinfo  [0]{\@secondoftwo}%
\providecommand \bibfield  [0]{\@secondoftwo}%
\providecommand \translation [1]{[#1]}%
\providecommand \BibitemOpen [0]{}%
\providecommand \bibitemStop [0]{}%
\providecommand \bibitemNoStop [0]{.\EOS\space}%
\providecommand \EOS [0]{\spacefactor3000\relax}%
\providecommand \BibitemShut  [1]{\csname bibitem#1\endcsname}%
\let\auto@bib@innerbib\@empty
\bibitem [{\citenamefont {Peccei}\ and\ \citenamefont {Quinn}(1977)}]{pq1977}%
  \BibitemOpen
  \bibfield  {author} {\bibinfo {author} {\bibfnamefont {R.~D.}\ \bibnamefont
  {Peccei}} and\ \bibinfo {author} {\bibfnamefont {H.~R.}\ \bibnamefont
  {Quinn}},\ }\bibfield  {title} {\enquote {\bibinfo {title} {{CP Conservation
  in the Presence of Pseudoparticles}},}\ }\href {\doibase
  10.1103/PhysRevLett.38.1440} {\bibfield  {journal} {\bibinfo  {journal}
  {Phys. Rev. Lett.}\ }\textbf {\bibinfo {volume} {38}},\ \bibinfo {pages}
  {1440} (\bibinfo {year} {1977})}\BibitemShut {NoStop}%
\bibitem [{\citenamefont {{Weinberg}}(1978)}]{Weinberg1978}%
  \BibitemOpen
  \bibfield  {author} {\bibinfo {author} {\bibfnamefont {S.}~\bibnamefont
  {{Weinberg}}},\ }\bibfield  {title} {\enquote {\bibinfo {title} {{A new light
  boson?}}}\ }\href {\doibase 10.1103/PhysRevLett.40.223} {\bibfield  {journal}
  {\bibinfo  {journal} {Physical Review Letters}\ }\textbf {\bibinfo {volume}
  {40}},\ \bibinfo {pages} {223} (\bibinfo {year} {1978})}\BibitemShut
  {NoStop}%
\bibitem [{\citenamefont {{Vysotsskii}}\ \emph {{\it
  et~al.}}(1978)\citenamefont {{Vysotsskii}}, \citenamefont {{Zel'dovich}},
  \citenamefont {{Khlopov}},\ and\ \citenamefont {{Chechetkin}}}]{axion78}%
  \BibitemOpen
  \bibfield  {author} {\bibinfo {author} {\bibfnamefont {M.~I.}\ \bibnamefont
  {{Vysotsskii}}}, \bibinfo {author} {\bibfnamefont {Y.~B.}\ \bibnamefont
  {{Zel'dovich}}}, \bibinfo {author} {\bibfnamefont {M.~Y.}\ \bibnamefont
  {{Khlopov}}}, and\ \bibinfo {author} {\bibfnamefont {V.~M.}\ \bibnamefont
  {{Chechetkin}}},\ }\bibfield  {title} {\enquote {\bibinfo {title} {{Some
  astrophysical limitations on the axion mass}},}\ }\href@noop {} {\bibfield
  {journal} {\bibinfo  {journal} {Soviet Journal of Experimental and
  Theoretical Physics Letters}\ }\textbf {\bibinfo {volume} {27}},\ \bibinfo
  {pages} {502} (\bibinfo {year} {1978})}\BibitemShut {NoStop}%
\bibitem [{\citenamefont {{Berezhiani}}\ \emph {{\it
  et~al.}}(1990)\citenamefont {{Berezhiani}}, \citenamefont {{Khlopov}},\ and\
  \citenamefont {{Khomeriki}}}]{axion90}%
  \BibitemOpen
  \bibfield  {author} {\bibinfo {author} {\bibfnamefont {Z.~G.}\ \bibnamefont
  {{Berezhiani}}}, \bibinfo {author} {\bibfnamefont {M.~Y.}\ \bibnamefont
  {{Khlopov}}}, and\ \bibinfo {author} {\bibfnamefont {R.~R.}\ \bibnamefont
  {{Khomeriki}}},\ }\bibfield  {title} {\enquote {\bibinfo {title} {{Nonthermal
  electromagnetic background from axion decays in the Universe in models with a
  low scale of family-symmetry breaking.}}}\ }\href@noop {} {\bibfield
  {journal} {\bibinfo  {journal} {Soviet Journal of Nuclear Physics}\ }\textbf
  {\bibinfo {volume} {52}},\ \bibinfo {pages} {65} (\bibinfo {year}
  {1990})}\BibitemShut {NoStop}%
\bibitem [{\citenamefont {{Anastassopoulos}}\ \emph {{\it
  et~al.}}(2017)\citenamefont {{Anastassopoulos}}, \citenamefont {{Aune}},
  \citenamefont {{Barth}}, \citenamefont {{Belov}}, \citenamefont
  {{Br{\"a}uninger}}, \citenamefont {{Cantatore}}, \citenamefont {{Carmona}},
  \citenamefont {{Castel}}, \citenamefont {{Cetin}}, \citenamefont
  {{Christensen}}, \citenamefont {{Collar}}, \citenamefont {{Dafni}},
  \citenamefont {{Davenport}}, \citenamefont {{Decker}}, \citenamefont
  {{Dermenev}}, \citenamefont {{Desch}}, \citenamefont {{Eleftheriadis}},
  \citenamefont {{Fanourakis}}, \citenamefont {{Ferrer-Ribas}}, \citenamefont
  {{Fischer}}, \citenamefont {{Garc{\'{\i}}a}}, \citenamefont {{Gardikiotis}},
  \citenamefont {{Garza}}, \citenamefont {{Gazis}}, \citenamefont {{Geralis}},
  \citenamefont {{Giomataris}}, \citenamefont {{Gninenko}}, \citenamefont
  {{Hailey}}, \citenamefont {{Hasinoff}}, \citenamefont {{Hoffmann}},
  \citenamefont {{Iguaz}}, \citenamefont {{Irastorza}}, \citenamefont
  {{Jakobsen}}, \citenamefont {{Jacoby}}, \citenamefont {{Jakov{\v c}i{\'c}}},
  \citenamefont {{Kaminski}}, \citenamefont {{Karuza}}, \citenamefont
  {{Kralj}}, \citenamefont {{Kr{\v c}mar}}, \citenamefont {{Kostoglou}},
  \citenamefont {{Krieger}}, \citenamefont {{Laki{\'c}}}, \citenamefont
  {{Laurent}}, \citenamefont {{Liolios}}, \citenamefont {{Ljubi{\v c}i{\'c}}},
  \citenamefont {{Luz{\'o}n}}, \citenamefont {{Maroudas}}, \citenamefont
  {{Miceli}}, \citenamefont {{Neff}}, \citenamefont {{Ortega}}, \citenamefont
  {{Papaevangelou}}, \citenamefont {{Paraschou}}, \citenamefont {{Pivovaroff}},
  \citenamefont {{Raffelt}}, \citenamefont {{Rosu}}, \citenamefont {{Ruz}},
  \citenamefont {{Ch{\'o}liz}}, \citenamefont {{Savvidis}}, \citenamefont
  {{Schmidt}}, \citenamefont {{Semertzidis}}, \citenamefont {{Solanki}},
  \citenamefont {{Stewart}}, \citenamefont {{Vafeiadis}}, \citenamefont
  {{Vogel}}, \citenamefont {{Yildiz}},\ and\ \citenamefont {{Zioutas}}}]{CAST}%
  \BibitemOpen
  \bibfield  {author} {\bibinfo {author} {\bibfnamefont {V.}~\bibnamefont
  {{Anastassopoulos}}} {\it et~al.},\ }\bibfield  {title} {\enquote {\bibinfo
  {title} {{New CAST limit on the axion-photon interaction}},}\ }\href
  {\doibase 10.1038/nphys4109} {\bibfield  {journal} {\bibinfo  {journal}
  {Nature Physics}\ }\textbf {\bibinfo {volume} {13}},\ \bibinfo {pages} {584}
  (\bibinfo {year} {2017})}\BibitemShut {NoStop}%
\bibitem [{\citenamefont {Asztalos}\ \emph {{\it et~al.}}(2004)\citenamefont
  {Asztalos}, \citenamefont {Bradley}, \citenamefont {Duffy}, \citenamefont
  {Hagmann}, \citenamefont {Kinion}, \citenamefont {Moltz}, \citenamefont
  {Rosenberg}, \citenamefont {Sikivie}, \citenamefont {Stoeffl}, \citenamefont
  {Sullivan}, \citenamefont {Tanner}, \citenamefont {van Bibber},\ and\
  \citenamefont {Yu}}]{alpex}%
  \BibitemOpen
  \bibfield  {author} {\bibinfo {author} {\bibfnamefont {S.~J.}\ \bibnamefont
  {Asztalos}} {\it et~al.},\ }\bibfield  {title} {\enquote {\bibinfo {title}
  {Improved rf cavity search for halo axions},}\ }\href {\doibase
  10.1103/PhysRevD.69.011101} {\bibfield  {journal} {\bibinfo  {journal} {Phys.
  Rev. D}\ }\textbf {\bibinfo {volume} {69}},\ \bibinfo {pages} {011101}
  (\bibinfo {year} {2004})}\BibitemShut {NoStop}%
\bibitem [{\citenamefont {Aprile}\ \emph {{\it et~al.}}(2017)\citenamefont
  {Aprile}, \citenamefont {Agostini}, \citenamefont {Alfonsi}, \citenamefont
  {Arisaka}, \citenamefont {Arneodo}, \citenamefont {Auger}, \citenamefont
  {Balan}, \citenamefont {Barrow}, \citenamefont {Baudis}, \citenamefont
  {Bauermeister}, \citenamefont {Behrens}, \citenamefont {Beltrame},
  \citenamefont {Bokeloh}, \citenamefont {Brown}, \citenamefont {Brown},
  \citenamefont {Bruenner}, \citenamefont {Bruno}, \citenamefont {Budnik},
  \citenamefont {Cardoso}, \citenamefont {Colijn}, \citenamefont {Contreras},
  \citenamefont {Cussonneau}, \citenamefont {Decowski}, \citenamefont
  {Duchovni}, \citenamefont {Fattori}, \citenamefont {Ferella}, \citenamefont
  {Fulgione}, \citenamefont {Gao}, \citenamefont {Garbini}, \citenamefont
  {Geis}, \citenamefont {Goetzke}, \citenamefont {Grignon}, \citenamefont
  {Gross}, \citenamefont {Hampel}, \citenamefont {Itay}, \citenamefont
  {Kaether}, \citenamefont {Kessler}, \citenamefont {Kish}, \citenamefont
  {Landsman}, \citenamefont {Lang}, \citenamefont {Le~Calloch}, \citenamefont
  {Lellouch}, \citenamefont {Levy}, \citenamefont {Lindemann}, \citenamefont
  {Lindner}, \citenamefont {Lopes}, \citenamefont {Lung}, \citenamefont
  {Lyashenko}, \citenamefont {MacMullin}, \citenamefont
  {Marrod\'an~Undagoitia}, \citenamefont {Masbou}, \citenamefont {Massoli},
  \citenamefont {Mayani~Paras}, \citenamefont {Melgarejo~Fernandez},
  \citenamefont {Meng}, \citenamefont {Messina}, \citenamefont {Miguez},
  \citenamefont {Molinario}, \citenamefont {Murra}, \citenamefont {Naganoma},
  \citenamefont {Ni}, \citenamefont {Oberlack}, \citenamefont {Orrigo},
  \citenamefont {Pantic}, \citenamefont {Persiani}, \citenamefont {Piastra},
  \citenamefont {Pienaar}, \citenamefont {Plante}, \citenamefont {Priel},
  \citenamefont {Reichard}, \citenamefont {Reuter}, \citenamefont {Rizzo},
  \citenamefont {Rosendahl}, \citenamefont {dos Santos}, \citenamefont
  {Sartorelli}, \citenamefont {Schindler}, \citenamefont {Schreiner},
  \citenamefont {Schumann}, \citenamefont {Scotto~Lavina}, \citenamefont
  {Selvi}, \citenamefont {Shagin}, \citenamefont {Simgen}, \citenamefont
  {Teymourian}, \citenamefont {Thers}, \citenamefont {Tiseni}, \citenamefont
  {Trinchero}, \citenamefont {Vitells}, \citenamefont {Wang}, \citenamefont
  {Weber},\ and\ \citenamefont {Weinheimer}}]{alpex1}%
  \BibitemOpen
  \bibfield  {author} {\bibinfo {author} {\bibfnamefont {E.}~\bibnamefont
  {Aprile}} {\it et~al.} (\bibinfo {collaboration} {XENON100 Collaboration}
  Collaboration),\ }\bibfield  {title} {\enquote {\bibinfo {title} {Erratum:
  First axion results from the xenon100 experiment [phys. rev. d 90, 062009
  (2014)]},}\ }\href {\doibase 10.1103/PhysRevD.95.029904} {\bibfield
  {journal} {\bibinfo  {journal} {Phys. Rev. D}\ }\textbf {\bibinfo {volume}
  {95}},\ \bibinfo {pages} {029904} (\bibinfo {year} {2017})}\BibitemShut
  {NoStop}%
\bibitem [{\citenamefont {Arias}\ \emph {{\it et~al.}}(2012)\citenamefont
  {Arias}, \citenamefont {Cadamuro}, \citenamefont {Goodsell}, \citenamefont
  {Jaeckel}, \citenamefont {Redondo},\ and\ \citenamefont {Ringwald}}]{ALPdm}%
  \BibitemOpen
  \bibfield  {author} {\bibinfo {author} {\bibfnamefont {P.}~\bibnamefont
  {Arias}}, \bibinfo {author} {\bibfnamefont {D.}~\bibnamefont {Cadamuro}},
  \bibinfo {author} {\bibfnamefont {M.}~\bibnamefont {Goodsell}}, \bibinfo
  {author} {\bibfnamefont {J.}~\bibnamefont {Jaeckel}}, \bibinfo {author}
  {\bibfnamefont {J.}~\bibnamefont {Redondo}}, and\ \bibinfo {author}
  {\bibfnamefont {A.}~\bibnamefont {Ringwald}},\ }\bibfield  {title} {\enquote
  {\bibinfo {title} {Wispy cold dark matter},}\ }\href
  {http://stacks.iop.org/1475-7516/2012/i=06/a=013} {\bibfield  {journal}
  {\bibinfo  {journal} {Journal of Cosmology and Astroparticle Physics}\
  }\textbf {\bibinfo {volume} {2012}},\ \bibinfo {pages} {013} (\bibinfo {year}
  {2012})}\BibitemShut {NoStop}%
\bibitem [{\citenamefont {{Hochmuth}}\ and\ \citenamefont
  {{Sigl}}(2007)}]{Hochmuth07}%
  \BibitemOpen
  \bibfield  {author} {\bibinfo {author} {\bibfnamefont {K.~A.}\ \bibnamefont
  {{Hochmuth}}} and\ \bibinfo {author} {\bibfnamefont {G.}~\bibnamefont
  {{Sigl}}},\ }\bibfield  {title} {\enquote {\bibinfo {title} {{Effects of
  axion-photon mixing on gamma-ray spectra from magnetized astrophysical
  sources}},}\ }\href {\doibase 10.1103/PhysRevD.76.123011} {\bibfield
  {journal} {\bibinfo  {journal} {\prd}\ }\textbf {\bibinfo {volume} {76}},\
  \bibinfo {eid} {123011} (\bibinfo {year} {2007})},\ \Eprint
  {http://arxiv.org/abs/0708.1144}{arXiv:0708.1144}\BibitemShut {NoStop}%
\bibitem [{\citenamefont {{Hooper}}\ and\ \citenamefont
  {{Serpico}}(2007)}]{Hooper07}%
  \BibitemOpen
  \bibfield  {author} {\bibinfo {author} {\bibfnamefont {D.}~\bibnamefont
  {{Hooper}}} and\ \bibinfo {author} {\bibfnamefont {P.~D.}\ \bibnamefont
  {{Serpico}}},\ }\bibfield  {title} {\enquote {\bibinfo {title} {{Detecting
  Axionlike Particles with Gamma Ray Telescopes}},}\ }\href {\doibase
  10.1103/PhysRevLett.99.231102} {\bibfield  {journal} {\bibinfo  {journal}
  {Physical Review Letters}\ }\textbf {\bibinfo {volume} {99}},\ \bibinfo {eid}
  {231102} (\bibinfo {year} {2007})},\ \Eprint
  {http://arxiv.org/abs/0706.3203}{arXiv:0706.3203}\BibitemShut {NoStop}%
\bibitem [{\citenamefont {{de Angelis}}\ \emph {{\it
  et~al.}}(2008)\citenamefont {{de Angelis}}, \citenamefont {{Mansutti}},\ and\
  \citenamefont {{Roncadelli}}}]{Angelis08}%
  \BibitemOpen
  \bibfield  {author} {\bibinfo {author} {\bibfnamefont {A.}~\bibnamefont {{de
  Angelis}}}, \bibinfo {author} {\bibfnamefont {O.}~\bibnamefont {{Mansutti}}},
  and\ \bibinfo {author} {\bibfnamefont {M.}~\bibnamefont {{Roncadelli}}},\
  }\bibfield  {title} {\enquote {\bibinfo {title} {{Axion-like particles,
  cosmic magnetic fields and gamma-ray astrophysics}},}\ }\href {\doibase
  10.1016/j.physletb.2007.12.012} {\bibfield  {journal} {\bibinfo  {journal}
  {Physics Letters B}\ }\textbf {\bibinfo {volume} {659}},\ \bibinfo {pages}
  {847} (\bibinfo {year} {2008})},\ \Eprint
  {http://arxiv.org/abs/0707.2695}{arXiv:0707.2695}\BibitemShut {NoStop}%
\bibitem [{\citenamefont {{Simet}}\ \emph {{\it et~al.}}(2008)\citenamefont
  {{Simet}}, \citenamefont {{Hooper}},\ and\ \citenamefont
  {{Serpico}}}]{Simet08}%
  \BibitemOpen
  \bibfield  {author} {\bibinfo {author} {\bibfnamefont {M.}~\bibnamefont
  {{Simet}}}, \bibinfo {author} {\bibfnamefont {D.}~\bibnamefont {{Hooper}}},
  and\ \bibinfo {author} {\bibfnamefont {P.~D.}\ \bibnamefont {{Serpico}}},\
  }\bibfield  {title} {\enquote {\bibinfo {title} {{Milky Way as a
  kiloparsec-scale axionscope}},}\ }\href {\doibase 10.1103/PhysRevD.77.063001}
  {\bibfield  {journal} {\bibinfo  {journal} {\prd}\ }\textbf {\bibinfo
  {volume} {77}},\ \bibinfo {eid} {063001} (\bibinfo {year} {2008})},\ \Eprint
  {http://arxiv.org/abs/0712.2825}{arXiv:0712.2825}\BibitemShut {NoStop}%
\bibitem [{\citenamefont {{S{\'a}nchez-Conde}}\ \emph {{\it
  et~al.}}(2009)\citenamefont {{S{\'a}nchez-Conde}}, \citenamefont {{Paneque}},
  \citenamefont {{Bloom}}, \citenamefont {{Prada}},\ and\ \citenamefont
  {{Dom{\'{\i}}nguez}}}]{sc09}%
  \BibitemOpen
  \bibfield  {author} {\bibinfo {author} {\bibfnamefont {M.~A.}\ \bibnamefont
  {{S{\'a}nchez-Conde}}}, \bibinfo {author} {\bibfnamefont {D.}~\bibnamefont
  {{Paneque}}}, \bibinfo {author} {\bibfnamefont {E.}~\bibnamefont {{Bloom}}},
  \bibinfo {author} {\bibfnamefont {F.}~\bibnamefont {{Prada}}}, and\ \bibinfo
  {author} {\bibfnamefont {A.}~\bibnamefont {{Dom{\'{\i}}nguez}}},\ }\bibfield
  {title} {\enquote {\bibinfo {title} {{Hints of the existence of axionlike
  particles from the gamma-ray spectra of cosmological sources}},}\ }\href
  {\doibase 10.1103/PhysRevD.79.123511} {\bibfield  {journal} {\bibinfo
  {journal} {\prd}\ }\textbf {\bibinfo {volume} {79}},\ \bibinfo {eid} {123511}
  (\bibinfo {year} {2009})},\ \Eprint
  {http://arxiv.org/abs/0905.3270}{arXiv:0905.3270}\BibitemShut {NoStop}%
\bibitem [{\citenamefont {{Meyer}}\ \emph {{\it et~al.}}(2013)\citenamefont
  {{Meyer}}, \citenamefont {{Horns}},\ and\ \citenamefont {{Raue}}}]{Meyer13}%
  \BibitemOpen
  \bibfield  {author} {\bibinfo {author} {\bibfnamefont {M.}~\bibnamefont
  {{Meyer}}}, \bibinfo {author} {\bibfnamefont {D.}~\bibnamefont {{Horns}}},
  and\ \bibinfo {author} {\bibfnamefont {M.}~\bibnamefont {{Raue}}},\
  }\bibfield  {title} {\enquote {\bibinfo {title} {{First lower limits on the
  photon-axion-like particle coupling from very high energy gamma-ray
  observations}},}\ }\href {\doibase 10.1103/PhysRevD.87.035027} {\bibfield
  {journal} {\bibinfo  {journal} {\prd}\ }\textbf {\bibinfo {volume} {87}},\
  \bibinfo {eid} {035027} (\bibinfo {year} {2013})},\ \Eprint
  {http://arxiv.org/abs/1302.1208}{arXiv:1302.1208}\BibitemShut {NoStop}%
\bibitem [{\citenamefont {{Ayala}}\ \emph {{\it et~al.}}(2014)\citenamefont
  {{Ayala}}, \citenamefont {{Dom{\'{\i}}nguez}}, \citenamefont {{Giannotti}},
  \citenamefont {{Mirizzi}},\ and\ \citenamefont {{Straniero}}}]{Ayala14}%
  \BibitemOpen
  \bibfield  {author} {\bibinfo {author} {\bibfnamefont {A.}~\bibnamefont
  {{Ayala}}}, \bibinfo {author} {\bibfnamefont {I.}~\bibnamefont
  {{Dom{\'{\i}}nguez}}}, \bibinfo {author} {\bibfnamefont {M.}~\bibnamefont
  {{Giannotti}}}, \bibinfo {author} {\bibfnamefont {A.}~\bibnamefont
  {{Mirizzi}}}, and\ \bibinfo {author} {\bibfnamefont {O.}~\bibnamefont
  {{Straniero}}},\ }\bibfield  {title} {\enquote {\bibinfo {title} {{Revisiting
  the Bound on Axion-Photon Coupling from Globular Clusters}},}\ }\href
  {\doibase 10.1103/PhysRevLett.113.191302} {\bibfield  {journal} {\bibinfo
  {journal} {Physical Review Letters}\ }\textbf {\bibinfo {volume} {113}},\
  \bibinfo {eid} {191302} (\bibinfo {year} {2014})},\ \Eprint
  {http://arxiv.org/abs/1406.6053}{arXiv:1406.6053}\BibitemShut {NoStop}%
\bibitem [{\citenamefont {{Belikov}}\ \emph {{\it et~al.}}(2011)\citenamefont
  {{Belikov}}, \citenamefont {{Goodenough}},\ and\ \citenamefont
  {{Hooper}}}]{belikov11alp}%
  \BibitemOpen
  \bibfield  {author} {\bibinfo {author} {\bibfnamefont {A.~V.}\ \bibnamefont
  {{Belikov}}}, \bibinfo {author} {\bibfnamefont {L.}~\bibnamefont
  {{Goodenough}}}, and\ \bibinfo {author} {\bibfnamefont {D.}~\bibnamefont
  {{Hooper}}},\ }\bibfield  {title} {\enquote {\bibinfo {title} {{No
  indications of axionlike particles from Fermi}},}\ }\href {\doibase
  10.1103/PhysRevD.83.063005} {\bibfield  {journal} {\bibinfo  {journal}
  {\prd}\ }\textbf {\bibinfo {volume} {83}},\ \bibinfo {eid} {063005} (\bibinfo
  {year} {2011})},\ \Eprint
  {http://arxiv.org/abs/1007.4862}{arXiv:1007.4862}\BibitemShut {NoStop}%
\bibitem [{\citenamefont {{Abramowski}}\ \emph {{\it
  et~al.}}(2013)\citenamefont {{Abramowski}}, \citenamefont {{Acero}},
  \citenamefont {{Aharonian}}, \citenamefont {{Ait Benkhali}}, \citenamefont
  {{Akhperjanian}}, \citenamefont {{Ang{\"u}ner}}, \citenamefont {{Anton}},
  \citenamefont {{Balenderan}}, \citenamefont {{Balzer}}, \citenamefont
  {{Barnacka}},\ and\ \citenamefont {et~al.}}]{hess13pks2155}%
  \BibitemOpen
  \bibfield  {author} {\bibinfo {author} {\bibfnamefont {A.}~\bibnamefont
  {{Abramowski}}} {\it et~al.},\ }\bibfield  {title} {\enquote {\bibinfo
  {title} {{Constraints on axionlike particles with H.E.S.S. from the
  irregularity of the PKS 2155-304 energy spectrum}},}\ }\href {\doibase
  10.1103/PhysRevD.88.102003} {\bibfield  {journal} {\bibinfo  {journal}
  {\prd}\ }\textbf {\bibinfo {volume} {88}},\ \bibinfo {eid} {102003} (\bibinfo
  {year} {2013})},\ \Eprint
  {http://arxiv.org/abs/1311.3148}{arXiv:1311.3148}\BibitemShut {NoStop}%
\bibitem [{\citenamefont {{Reesman}}\ and\ \citenamefont
  {{Walker}}(2014)}]{reesman14alp}%
  \BibitemOpen
  \bibfield  {author} {\bibinfo {author} {\bibfnamefont {R.}~\bibnamefont
  {{Reesman}}} and\ \bibinfo {author} {\bibfnamefont {T.~P.}\ \bibnamefont
  {{Walker}}},\ }\bibfield  {title} {\enquote {\bibinfo {title} {{Probing the
  scale of ALP interactions with Fermi blazars}},}\ }\href {\doibase
  10.1088/1475-7516/2014/08/021} {\bibfield  {journal} {\bibinfo  {journal}
  {\jcap}\ }\textbf {\bibinfo {volume} {8}},\ \bibinfo {eid} {021} (\bibinfo
  {year} {2014})},\ \Eprint
  {http://arxiv.org/abs/1402.2533}{arXiv:1402.2533}\BibitemShut {NoStop}%
\bibitem [{\citenamefont {{Berenji}}\ \emph {{\it et~al.}}(2016)\citenamefont
  {{Berenji}}, \citenamefont {{Gaskins}},\ and\ \citenamefont
  {{Meyer}}}]{berenji16alp_ns}%
  \BibitemOpen
  \bibfield  {author} {\bibinfo {author} {\bibfnamefont {B.}~\bibnamefont
  {{Berenji}}}, \bibinfo {author} {\bibfnamefont {J.}~\bibnamefont
  {{Gaskins}}}, and\ \bibinfo {author} {\bibfnamefont {M.}~\bibnamefont
  {{Meyer}}},\ }\bibfield  {title} {\enquote {\bibinfo {title} {{Constraints on
  axions and axionlike particles from Fermi Large Area Telescope observations
  of neutron stars}},}\ }\href {\doibase 10.1103/PhysRevD.93.045019} {\bibfield
   {journal} {\bibinfo  {journal} {\prd}\ }\textbf {\bibinfo {volume} {93}},\
  \bibinfo {eid} {045019} (\bibinfo {year} {2016})},\ \Eprint
  {http://arxiv.org/abs/1602.00091}{arXiv:1602.00091}\BibitemShut {NoStop}%
\bibitem [{\citenamefont {{Ajello}}\ \emph {{\it et~al.}}(2016)\citenamefont
  {{Ajello}}, \citenamefont {{Albert}}, \citenamefont {{Anderson}},
  \citenamefont {{Baldini}}, \citenamefont {{Barbiellini}}, \citenamefont
  {{Bastieri}}, \citenamefont {{Bellazzini}}, \citenamefont {{Bissaldi}},
  \citenamefont {{Blandford}}, \citenamefont {{Bloom}}, \citenamefont
  {{Bonino}}, \citenamefont {{Bottacini}}, \citenamefont {{Bregeon}},
  \citenamefont {{Bruel}}, \citenamefont {{Buehler}}, \citenamefont
  {{Caliandro}}, \citenamefont {{Cameron}}, \citenamefont {{Caragiulo}},
  \citenamefont {{Caraveo}}, \citenamefont {{Cecchi}}, \citenamefont
  {{Chekhtman}}, \citenamefont {{Ciprini}}, \citenamefont {{Cohen-Tanugi}},
  \citenamefont {{Conrad}}, \citenamefont {{Costanza}}, \citenamefont
  {{D'Ammando}}, \citenamefont {{de Angelis}}, \citenamefont {{de Palma}},
  \citenamefont {{Desiante}}, \citenamefont {{Di Mauro}}, \citenamefont {{Di
  Venere}}, \citenamefont {{Dom{\'{\i}}nguez}}, \citenamefont {{Drell}},
  \citenamefont {{Favuzzi}}, \citenamefont {{Focke}}, \citenamefont
  {{Franckowiak}}, \citenamefont {{Fukazawa}}, \citenamefont {{Funk}},
  \citenamefont {{Fusco}}, \citenamefont {{Gargano}}, \citenamefont
  {{Gasparrini}}, \citenamefont {{Giglietto}}, \citenamefont {{Glanzman}},
  \citenamefont {{Godfrey}}, \citenamefont {{Guiriec}}, \citenamefont
  {{Horan}}, \citenamefont {{J{\'o}hannesson}}, \citenamefont {{Katsuragawa}},
  \citenamefont {{Kensei}}, \citenamefont {{Kuss}}, \citenamefont {{Larsson}},
  \citenamefont {{Latronico}}, \citenamefont {{Li}}, \citenamefont {{Li}},
  \citenamefont {{Longo}}, \citenamefont {{Loparco}}, \citenamefont
  {{Lubrano}}, \citenamefont {{Madejski}}, \citenamefont {{Maldera}},
  \citenamefont {{Manfreda}}, \citenamefont {{Mayer}}, \citenamefont
  {{Mazziotta}}, \citenamefont {{Meyer}}, \citenamefont {{Michelson}},
  \citenamefont {{Mirabal}}, \citenamefont {{Mizuno}}, \citenamefont
  {{Monzani}}, \citenamefont {{Morselli}}, \citenamefont {{Moskalenko}},
  \citenamefont {{Murgia}}, \citenamefont {{Negro}}, \citenamefont {{Nuss}},
  \citenamefont {{Okada}}, \citenamefont {{Orlando}}, \citenamefont {{Ormes}},
  \citenamefont {{Paneque}}, \citenamefont {{Perkins}}, \citenamefont
  {{Pesce-Rollins}}, \citenamefont {{Piron}}, \citenamefont {{Pivato}},
  \citenamefont {{Porter}}, \citenamefont {{Rain{\`o}}}, \citenamefont
  {{Rando}}, \citenamefont {{Razzano}}, \citenamefont {{Reimer}}, \citenamefont
  {{S{\'a}nchez-Conde}}, \citenamefont {{Sgr{\`o}}}, \citenamefont {{Simone}},
  \citenamefont {{Siskind}}, \citenamefont {{Spada}}, \citenamefont
  {{Spandre}}, \citenamefont {{Spinelli}}, \citenamefont {{Takahashi}},
  \citenamefont {{Thayer}}, \citenamefont {{Torres}}, \citenamefont {{Tosti}},
  \citenamefont {{Troja}}, \citenamefont {{Uchiyama}}, \citenamefont {{Wood}},
  \citenamefont {{Wood}}, \citenamefont {{Zaharijas}}, \citenamefont
  {{Zimmer}},\ and\ \citenamefont {{Fermi-LAT Collaboration}}}]{fermi16alp}%
  \BibitemOpen
  \bibfield  {author} {\bibinfo {author} {\bibfnamefont {M.}~\bibnamefont
  {{Ajello}}} {\it et~al.},\ }\bibfield  {title} {\enquote {\bibinfo {title}
  {{Search for Spectral Irregularities due to Photon-Axionlike-Particle
  Oscillations with the Fermi Large Area Telescope}},}\ }\href {\doibase
  10.1103/PhysRevLett.116.161101} {\bibfield  {journal} {\bibinfo  {journal}
  {Physical Review Letters}\ }\textbf {\bibinfo {volume} {116}},\ \bibinfo
  {eid} {161101} (\bibinfo {year} {2016})},\ \Eprint
  {http://arxiv.org/abs/1603.06978}{arXiv:1603.06978}\BibitemShut {NoStop}%
\bibitem [{\citenamefont {{Meyer}}\ \emph {{\it et~al.}}(2017)\citenamefont
  {{Meyer}}, \citenamefont {{Giannotti}}, \citenamefont {{Mirizzi}},
  \citenamefont {{Conrad}},\ and\ \citenamefont
  {{S{\'a}nchez-Conde}}}]{meyer17sne}%
  \BibitemOpen
  \bibfield  {author} {\bibinfo {author} {\bibfnamefont {M.}~\bibnamefont
  {{Meyer}}}, \bibinfo {author} {\bibfnamefont {M.}~\bibnamefont
  {{Giannotti}}}, \bibinfo {author} {\bibfnamefont {A.}~\bibnamefont
  {{Mirizzi}}}, \bibinfo {author} {\bibfnamefont {J.}~\bibnamefont {{Conrad}}},
  and\ \bibinfo {author} {\bibfnamefont {M.~A.}\ \bibnamefont
  {{S{\'a}nchez-Conde}}},\ }\bibfield  {title} {\enquote {\bibinfo {title}
  {{Fermi Large Area Telescope as a Galactic Supernovae Axionscope}},}\ }\href
  {\doibase 10.1103/PhysRevLett.118.011103} {\bibfield  {journal} {\bibinfo
  {journal} {Physical Review Letters}\ }\textbf {\bibinfo {volume} {118}},\
  \bibinfo {eid} {011103} (\bibinfo {year} {2017})},\ \Eprint
  {http://arxiv.org/abs/1609.02350}{arXiv:1609.02350}\BibitemShut {NoStop}%
\bibitem [{\citenamefont {{Zhang}}\ \emph {{\it et~al.}}(2018)\citenamefont
  {{Zhang}}, \citenamefont {{Liang}}, \citenamefont {{Li}}, \citenamefont
  {{Liao}}, \citenamefont {{Feng}}, \citenamefont {{Yuan}}, \citenamefont
  {{Fan}},\ and\ \citenamefont {{Ren}}}]{zc16alp}%
  \BibitemOpen
  \bibfield  {author} {\bibinfo {author} {\bibfnamefont {C.}~\bibnamefont
  {{Zhang}}}, \bibinfo {author} {\bibfnamefont {Y.-F.}\ \bibnamefont
  {{Liang}}}, \bibinfo {author} {\bibfnamefont {S.}~\bibnamefont {{Li}}},
  \bibinfo {author} {\bibfnamefont {N.-H.}\ \bibnamefont {{Liao}}}, \bibinfo
  {author} {\bibfnamefont {L.}~\bibnamefont {{Feng}}}, \bibinfo {author}
  {\bibfnamefont {Q.}~\bibnamefont {{Yuan}}}, \bibinfo {author} {\bibfnamefont
  {Y.-Z.}\ \bibnamefont {{Fan}}}, and\ \bibinfo {author} {\bibfnamefont
  {Z.-Z.}\ \bibnamefont {{Ren}}},\ }\bibfield  {title} {\enquote {\bibinfo
  {title} {{New Bounds on Axion-Like Particles From the Fermi Large Area
  Telescope observation of PKS $2155-304$}},}\ }\href@noop {} {\bibfield
  {journal} {\bibinfo  {journal} {ArXiv e-prints}\ } (\bibinfo {year}
  {2018})},\ \Eprint
  {http://arxiv.org/abs/1802.08420}{arXiv:1802.08420}\BibitemShut {NoStop}%
\bibitem [{\citenamefont {{Majumdar}}\ \emph {{\it
  et~al.}}(2017{\natexlab{a}})\citenamefont {{Majumdar}}, \citenamefont
  {{Calore}},\ and\ \citenamefont {{Horns}}}]{Majumdar17psr1}%
  \BibitemOpen
  \bibfield  {author} {\bibinfo {author} {\bibfnamefont {J.}~\bibnamefont
  {{Majumdar}}}, \bibinfo {author} {\bibfnamefont {F.}~\bibnamefont
  {{Calore}}}, and\ \bibinfo {author} {\bibfnamefont {D.}~\bibnamefont
  {{Horns}}},\ }\bibfield  {title} {\enquote {\bibinfo {title} {{Modulations in
  Spectra of Galactic Gamma-ray sources as a result of photon-ALPs mixing}},}\
  }\href@noop {} {\bibfield  {journal} {\bibinfo  {journal} {ArXiv e-prints}\ }
  (\bibinfo {year} {2017}{\natexlab{a}})},\ \Eprint
  {http://arxiv.org/abs/1710.09894}{arXiv:1710.09894}\BibitemShut {NoStop}%
\bibitem [{\citenamefont {{Majumdar}}\ \emph {{\it et~al.}}(2018)\citenamefont
  {{Majumdar}}, \citenamefont {{Calore}},\ and\ \citenamefont
  {{Horns}}}]{ALP2018}%
  \BibitemOpen
  \bibfield  {author} {\bibinfo {author} {\bibfnamefont {J.}~\bibnamefont
  {{Majumdar}}}, \bibinfo {author} {\bibfnamefont {F.}~\bibnamefont
  {{Calore}}}, and\ \bibinfo {author} {\bibfnamefont {D.}~\bibnamefont
  {{Horns}}},\ }\bibfield  {title} {\enquote {\bibinfo {title} {{Gamma-ray
  spectral modulations of Galactic pulsars caused by photon-ALPs mixing}},}\
  }\href@noop {} {\bibfield  {journal} {\bibinfo  {journal} {ArXiv e-prints}\ }
  (\bibinfo {year} {2018})},\ \Eprint
  {http://arxiv.org/abs/1801.08813}{arXiv:1801.08813}\BibitemShut {NoStop}%
\bibitem [{\citenamefont {{Majumdar}}\ \emph {{\it
  et~al.}}(2017{\natexlab{b}})\citenamefont {{Majumdar}}, \citenamefont
  {{Calore}},\ and\ \citenamefont {{Horns}}}]{Majumdar17psr2}%
  \BibitemOpen
  \bibfield  {author} {\bibinfo {author} {\bibfnamefont {J.}~\bibnamefont
  {{Majumdar}}}, \bibinfo {author} {\bibfnamefont {F.}~\bibnamefont
  {{Calore}}}, and\ \bibinfo {author} {\bibfnamefont {D.}~\bibnamefont
  {{Horns}}},\ }\bibfield  {title} {\enquote {\bibinfo {title} {{Spectral
  modulation of non-Galactic plane Gamma-ray pulsars due to photon-ALPs mixing
  in Galactic magnetic field}},}\ }\href@noop {} {\bibfield  {journal}
  {\bibinfo  {journal} {ArXiv e-prints}\ } (\bibinfo {year}
  {2017}{\natexlab{b}})},\ \Eprint
  {http://arxiv.org/abs/1711.08723}{arXiv:1711.08723}\BibitemShut {NoStop}%
\bibitem [{\citenamefont {{Vogel}}\ \emph {{\it et~al.}}(2017)\citenamefont
  {{Vogel}}, \citenamefont {{Laha}},\ and\ \citenamefont
  {{Meyer}}}]{ALP2017LHAASO}%
  \BibitemOpen
  \bibfield  {author} {\bibinfo {author} {\bibfnamefont {H.}~\bibnamefont
  {{Vogel}}}, \bibinfo {author} {\bibfnamefont {R.}~\bibnamefont {{Laha}}},
  and\ \bibinfo {author} {\bibfnamefont {M.}~\bibnamefont {{Meyer}}},\
  }\bibfield  {title} {\enquote {\bibinfo {title} {{Diffuse axion-like particle
  searches}},}\ }\href@noop {} {\bibfield  {journal} {\bibinfo  {journal}
  {ArXiv e-prints}\ } (\bibinfo {year} {2017})},\ \Eprint
  {http://arxiv.org/abs/1712.01839}{arXiv:1712.01839}\BibitemShut {NoStop}%
\bibitem [{\citenamefont {{Acero}}\ \emph {{\it et~al.}}(2015)\citenamefont
  {{Acero}}, \citenamefont {{Ackermann}}, \citenamefont {{Ajello}},
  \citenamefont {{Albert}}, \citenamefont {{Atwood}}, \citenamefont
  {{Axelsson}}, \citenamefont {{Baldini}}, \citenamefont {{Ballet}},
  \citenamefont {{Barbiellini}}, \citenamefont {{Bastieri}}, \citenamefont
  {{Belfiore}}, \citenamefont {{Bellazzini}}, \citenamefont {{Bissaldi}},
  \citenamefont {{Blandford}}, \citenamefont {{Bloom}}, \citenamefont
  {{Bogart}}, \citenamefont {{Bonino}}, \citenamefont {{Bottacini}},
  \citenamefont {{Bregeon}}, \citenamefont {{Britto}}, \citenamefont {{Bruel}},
  \citenamefont {{Buehler}}, \citenamefont {{Burnett}}, \citenamefont
  {{Buson}}, \citenamefont {{Caliandro}}, \citenamefont {{Cameron}},
  \citenamefont {{Caputo}}, \citenamefont {{Caragiulo}}, \citenamefont
  {{Caraveo}}, \citenamefont {{Casandjian}}, \citenamefont {{Cavazzuti}},
  \citenamefont {{Charles}}, \citenamefont {{Chaves}}, \citenamefont
  {{Chekhtman}}, \citenamefont {{Cheung}}, \citenamefont {{Chiang}},
  \citenamefont {{Chiaro}}, \citenamefont {{Ciprini}}, \citenamefont {{Claus}},
  \citenamefont {{Cohen-Tanugi}}, \citenamefont {{Cominsky}}, \citenamefont
  {{Conrad}}, \citenamefont {{Cutini}}, \citenamefont {{D'Ammando}},
  \citenamefont {{de Angelis}}, \citenamefont {{DeKlotz}}, \citenamefont {{de
  Palma}}, \citenamefont {{Desiante}}, \citenamefont {{Digel}}, \citenamefont
  {{Di Venere}}, \citenamefont {{Drell}}, \citenamefont {{Dubois}},
  \citenamefont {{Dumora}}, \citenamefont {{Favuzzi}}, \citenamefont {{Fegan}},
  \citenamefont {{Ferrara}}, \citenamefont {{Finke}}, \citenamefont
  {{Franckowiak}}, \citenamefont {{Fukazawa}}, \citenamefont {{Funk}},
  \citenamefont {{Fusco}}, \citenamefont {{Gargano}}, \citenamefont
  {{Gasparrini}}, \citenamefont {{Giebels}}, \citenamefont {{Giglietto}},
  \citenamefont {{Giommi}}, \citenamefont {{Giordano}}, \citenamefont
  {{Giroletti}}, \citenamefont {{Glanzman}}, \citenamefont {{Godfrey}},
  \citenamefont {{Grenier}}, \citenamefont {{Grondin}}, \citenamefont
  {{Grove}}, \citenamefont {{Guillemot}}, \citenamefont {{Guiriec}},
  \citenamefont {{Hadasch}}, \citenamefont {{Harding}}, \citenamefont {{Hays}},
  \citenamefont {{Hewitt}}, \citenamefont {{Hill}}, \citenamefont {{Horan}},
  \citenamefont {{Iafrate}}, \citenamefont {{Jogler}}, \citenamefont
  {{J{\'o}hannesson}}, \citenamefont {{Johnson}}, \citenamefont {{Johnson}},
  \citenamefont {{Johnson}}, \citenamefont {{Johnson}}, \citenamefont
  {{Kamae}}, \citenamefont {{Kataoka}}, \citenamefont {{Katsuta}},
  \citenamefont {{Kuss}}, \citenamefont {{La Mura}}, \citenamefont {{Landriu}},
  \citenamefont {{Larsson}}, \citenamefont {{Latronico}}, \citenamefont
  {{Lemoine-Goumard}}, \citenamefont {{Li}}, \citenamefont {{Li}},
  \citenamefont {{Longo}}, \citenamefont {{Loparco}}, \citenamefont {{Lott}},
  \citenamefont {{Lovellette}}, \citenamefont {{Lubrano}}, \citenamefont
  {{Madejski}}, \citenamefont {{Massaro}}, \citenamefont {{Mayer}},
  \citenamefont {{Mazziotta}}, \citenamefont {{McEnery}}, \citenamefont
  {{Michelson}}, \citenamefont {{Mirabal}}, \citenamefont {{Mizuno}},
  \citenamefont {{Moiseev}}, \citenamefont {{Mongelli}}, \citenamefont
  {{Monzani}}, \citenamefont {{Morselli}}, \citenamefont {{Moskalenko}},
  \citenamefont {{Murgia}}, \citenamefont {{Nuss}}, \citenamefont {{Ohno}},
  \citenamefont {{Ohsugi}}, \citenamefont {{Omodei}}, \citenamefont
  {{Orienti}}, \citenamefont {{Orlando}}, \citenamefont {{Ormes}},
  \citenamefont {{Paneque}}, \citenamefont {{Panetta}}, \citenamefont
  {{Perkins}}, \citenamefont {{Pesce-Rollins}}, \citenamefont {{Piron}},
  \citenamefont {{Pivato}}, \citenamefont {{Porter}}, \citenamefont
  {{Racusin}}, \citenamefont {{Rando}}, \citenamefont {{Razzano}},
  \citenamefont {{Razzaque}}, \citenamefont {{Reimer}}, \citenamefont
  {{Reimer}}, \citenamefont {{Reposeur}}, \citenamefont {{Rochester}},
  \citenamefont {{Romani}}, \citenamefont {{Salvetti}}, \citenamefont
  {{S{\'a}nchez-Conde}}, \citenamefont {{Saz Parkinson}}, \citenamefont
  {{Schulz}}, \citenamefont {{Siskind}}, \citenamefont {{Smith}}, \citenamefont
  {{Spada}}, \citenamefont {{Spandre}}, \citenamefont {{Spinelli}},
  \citenamefont {{Stephens}}, \citenamefont {{Strong}}, \citenamefont
  {{Suson}}, \citenamefont {{Takahashi}}, \citenamefont {{Takahashi}},
  \citenamefont {{Tanaka}}, \citenamefont {{Thayer}}, \citenamefont {{Thayer}},
  \citenamefont {{Thompson}}, \citenamefont {{Tibaldo}}, \citenamefont
  {{Tibolla}}, \citenamefont {{Torres}}, \citenamefont {{Torresi}},
  \citenamefont {{Tosti}}, \citenamefont {{Troja}}, \citenamefont {{Van
  Klaveren}}, \citenamefont {{Vianello}}, \citenamefont {{Winer}},
  \citenamefont {{Wood}}, \citenamefont {{Wood}}, \citenamefont {{Zimmer}},\
  and\ \citenamefont {{Fermi-LAT Collaboration}}}]{3fgl}%
  \BibitemOpen
  \bibfield  {author} {\bibinfo {author} {\bibfnamefont {F.}~\bibnamefont
  {{Acero}}} {\it et~al.},\ }\bibfield  {title} {\enquote {\bibinfo {title}
  {{Fermi Large Area Telescope Third Source Catalog}},}\ }\href {\doibase
  10.1088/0067-0049/218/2/23} {\bibfield  {journal} {\bibinfo  {journal}
  {Astrophys. J. Suppl.}\ }\textbf {\bibinfo {volume} {218}},\ \bibinfo {eid}
  {23} (\bibinfo {year} {2015})},\ \Eprint
  {http://arxiv.org/abs/1501.02003}{arXiv:1501.02003}\BibitemShut {NoStop}%
\bibitem [{\citenamefont {{Acero}}\ \emph {{\it et~al.}}(2016)\citenamefont
  {{Acero}}, \citenamefont {{Ackermann}}, \citenamefont {{Ajello}},
  \citenamefont {{Baldini}}, \citenamefont {{Ballet}}, \citenamefont
  {{Barbiellini}}, \citenamefont {{Bastieri}}, \citenamefont {{Bellazzini}},
  \citenamefont {{Bissaldi}}, \citenamefont {{Blandford}}, \citenamefont
  {{Bloom}}, \citenamefont {{Bonino}}, \citenamefont {{Bottacini}},
  \citenamefont {{Brandt}}, \citenamefont {{Bregeon}}, \citenamefont {{Bruel}},
  \citenamefont {{Buehler}}, \citenamefont {{Buson}}, \citenamefont
  {{Caliandro}}, \citenamefont {{Cameron}}, \citenamefont {{Caputo}},
  \citenamefont {{Caragiulo}}, \citenamefont {{Caraveo}}, \citenamefont
  {{Casandjian}}, \citenamefont {{Cavazzuti}}, \citenamefont {{Cecchi}},
  \citenamefont {{Chekhtman}}, \citenamefont {{Chiang}}, \citenamefont
  {{Chiaro}}, \citenamefont {{Ciprini}}, \citenamefont {{Claus}}, \citenamefont
  {{Cohen}}, \citenamefont {{Cohen-Tanugi}}, \citenamefont {{Cominsky}},
  \citenamefont {{Condon}}, \citenamefont {{Conrad}}, \citenamefont {{Cutini}},
  \citenamefont {{D'Ammando}}, \citenamefont {{de Angelis}}, \citenamefont {{de
  Palma}}, \citenamefont {{Desiante}}, \citenamefont {{Digel}}, \citenamefont
  {{Di Venere}}, \citenamefont {{Drell}}, \citenamefont {{Drlica-Wagner}},
  \citenamefont {{Favuzzi}}, \citenamefont {{Ferrara}}, \citenamefont
  {{Franckowiak}}, \citenamefont {{Fukazawa}}, \citenamefont {{Funk}},
  \citenamefont {{Fusco}}, \citenamefont {{Gargano}}, \citenamefont
  {{Gasparrini}}, \citenamefont {{Giglietto}}, \citenamefont {{Giommi}},
  \citenamefont {{Giordano}}, \citenamefont {{Giroletti}}, \citenamefont
  {{Glanzman}}, \citenamefont {{Godfrey}}, \citenamefont {{Gomez-Vargas}},
  \citenamefont {{Grenier}}, \citenamefont {{Grondin}}, \citenamefont
  {{Guillemot}}, \citenamefont {{Guiriec}}, \citenamefont {{Gustafsson}},
  \citenamefont {{Hadasch}}, \citenamefont {{Harding}}, \citenamefont
  {{Hayashida}}, \citenamefont {{Hays}}, \citenamefont {{Hewitt}},
  \citenamefont {{Hill}}, \citenamefont {{Horan}}, \citenamefont {{Hou}},
  \citenamefont {{Iafrate}}, \citenamefont {{Jogler}}, \citenamefont
  {{J{\'o}hannesson}}, \citenamefont {{Johnson}}, \citenamefont {{Kamae}},
  \citenamefont {{Katagiri}}, \citenamefont {{Kataoka}}, \citenamefont
  {{Katsuta}}, \citenamefont {{Kerr}}, \citenamefont {{Kn{\"o}dlseder}},
  \citenamefont {{Kocevski}}, \citenamefont {{Kuss}}, \citenamefont {{Laffon}},
  \citenamefont {{Lande}}, \citenamefont {{Larsson}}, \citenamefont
  {{Latronico}}, \citenamefont {{Lemoine-Goumard}}, \citenamefont {{Li}},
  \citenamefont {{Li}}, \citenamefont {{Longo}}, \citenamefont {{Loparco}},
  \citenamefont {{Lovellette}}, \citenamefont {{Lubrano}}, \citenamefont
  {{Magill}}, \citenamefont {{Maldera}}, \citenamefont {{Marelli}},
  \citenamefont {{Mayer}}, \citenamefont {{Mazziotta}}, \citenamefont
  {{Michelson}}, \citenamefont {{Mitthumsiri}}, \citenamefont {{Mizuno}},
  \citenamefont {{Moiseev}}, \citenamefont {{Monzani}}, \citenamefont
  {{Moretti}}, \citenamefont {{Morselli}}, \citenamefont {{Moskalenko}},
  \citenamefont {{Murgia}}, \citenamefont {{Nemmen}}, \citenamefont {{Nuss}},
  \citenamefont {{Ohsugi}}, \citenamefont {{Omodei}}, \citenamefont
  {{Orienti}}, \citenamefont {{Orlando}}, \citenamefont {{Ormes}},
  \citenamefont {{Paneque}}, \citenamefont {{Perkins}}, \citenamefont
  {{Pesce-Rollins}}, \citenamefont {{Petrosian}}, \citenamefont {{Piron}},
  \citenamefont {{Pivato}}, \citenamefont {{Porter}}, \citenamefont
  {{Rain{\`o}}}, \citenamefont {{Rando}}, \citenamefont {{Razzano}},
  \citenamefont {{Razzaque}}, \citenamefont {{Reimer}}, \citenamefont
  {{Reimer}}, \citenamefont {{Renaud}}, \citenamefont {{Reposeur}},
  \citenamefont {{Rousseau}}, \citenamefont {{Saz Parkinson}}, \citenamefont
  {{Schmid}}, \citenamefont {{Schulz}}, \citenamefont {{Sgr{\`o}}},
  \citenamefont {{Siskind}}, \citenamefont {{Spada}}, \citenamefont
  {{Spandre}}, \citenamefont {{Spinelli}}, \citenamefont {{Strong}},
  \citenamefont {{Suson}}, \citenamefont {{Tajima}}, \citenamefont
  {{Takahashi}}, \citenamefont {{Tanaka}}, \citenamefont {{Thayer}},
  \citenamefont {{Thompson}}, \citenamefont {{Tibaldo}}, \citenamefont
  {{Tibolla}}, \citenamefont {{Torres}}, \citenamefont {{Tosti}}, \citenamefont
  {{Troja}}, \citenamefont {{Uchiyama}}, \citenamefont {{Vianello}},
  \citenamefont {{Wells}}, \citenamefont {{Wood}}, \citenamefont {{Wood}},
  \citenamefont {{Yassine}}, \citenamefont {{den Hartog}},\ and\ \citenamefont
  {{Zimmer}}}]{fermi16snrcata}%
  \BibitemOpen
  \bibfield  {author} {\bibinfo {author} {\bibfnamefont {F.}~\bibnamefont
  {{Acero}}} {\it et~al.},\ }\bibfield  {title} {\enquote {\bibinfo {title}
  {{The First Fermi LAT Supernova Remnant Catalog}},}\ }\href {\doibase
  10.3847/0067-0049/224/1/8} {\bibfield  {journal} {\bibinfo  {journal}
  {Astrophys. J. Suppl.}\ }\textbf {\bibinfo {volume} {224}},\ \bibinfo {eid}
  {8} (\bibinfo {year} {2016})},\ \Eprint
  {http://arxiv.org/abs/1511.06778}{arXiv:1511.06778}\BibitemShut {NoStop}%
\bibitem [{\citenamefont {{Ackermann}}\ \emph {{\it et~al.}}(2013)\citenamefont
  {{Ackermann}}, \citenamefont {{Ajello}}, \citenamefont {{Allafort}},
  \citenamefont {{Baldini}}, \citenamefont {{Ballet}}, \citenamefont
  {{Barbiellini}}, \citenamefont {{Baring}}, \citenamefont {{Bastieri}},
  \citenamefont {{Bechtol}}, \citenamefont {{Bellazzini}}, \citenamefont
  {{Blandford}}, \citenamefont {{Bloom}}, \citenamefont {{Bonamente}},
  \citenamefont {{Borgland}}, \citenamefont {{Bottacini}}, \citenamefont
  {{Brandt}}, \citenamefont {{Bregeon}}, \citenamefont {{Brigida}},
  \citenamefont {{Bruel}}, \citenamefont {{Buehler}}, \citenamefont
  {{Busetto}}, \citenamefont {{Buson}}, \citenamefont {{Caliandro}},
  \citenamefont {{Cameron}}, \citenamefont {{Caraveo}}, \citenamefont
  {{Casandjian}}, \citenamefont {{Cecchi}}, \citenamefont {{{\c C}elik}},
  \citenamefont {{Charles}}, \citenamefont {{Chaty}}, \citenamefont {{Chaves}},
  \citenamefont {{Chekhtman}}, \citenamefont {{Cheung}}, \citenamefont
  {{Chiang}}, \citenamefont {{Chiaro}}, \citenamefont {{Cillis}}, \citenamefont
  {{Ciprini}}, \citenamefont {{Claus}}, \citenamefont {{Cohen-Tanugi}},
  \citenamefont {{Cominsky}}, \citenamefont {{Conrad}}, \citenamefont
  {{Corbel}}, \citenamefont {{Cutini}}, \citenamefont {{D'Ammando}},
  \citenamefont {{de Angelis}}, \citenamefont {{de Palma}}, \citenamefont
  {{Dermer}}, \citenamefont {{do Couto e Silva}}, \citenamefont {{Drell}},
  \citenamefont {{Drlica-Wagner}}, \citenamefont {{Falletti}}, \citenamefont
  {{Favuzzi}}, \citenamefont {{Ferrara}}, \citenamefont {{Franckowiak}},
  \citenamefont {{Fukazawa}}, \citenamefont {{Funk}}, \citenamefont {{Fusco}},
  \citenamefont {{Gargano}}, \citenamefont {{Germani}}, \citenamefont
  {{Giglietto}}, \citenamefont {{Giommi}}, \citenamefont {{Giordano}},
  \citenamefont {{Giroletti}}, \citenamefont {{Glanzman}}, \citenamefont
  {{Godfrey}}, \citenamefont {{Grenier}}, \citenamefont {{Grondin}},
  \citenamefont {{Grove}}, \citenamefont {{Guiriec}}, \citenamefont
  {{Hadasch}}, \citenamefont {{Hanabata}}, \citenamefont {{Harding}},
  \citenamefont {{Hayashida}}, \citenamefont {{Hayashi}}, \citenamefont
  {{Hays}}, \citenamefont {{Hewitt}}, \citenamefont {{Hill}}, \citenamefont
  {{Hughes}}, \citenamefont {{Jackson}}, \citenamefont {{Jogler}},
  \citenamefont {{J{\'o}hannesson}}, \citenamefont {{Johnson}}, \citenamefont
  {{Kamae}}, \citenamefont {{Kataoka}}, \citenamefont {{Katsuta}},
  \citenamefont {{Kn{\"o}dlseder}}, \citenamefont {{Kuss}}, \citenamefont
  {{Lande}}, \citenamefont {{Larsson}}, \citenamefont {{Latronico}},
  \citenamefont {{Lemoine-Goumard}}, \citenamefont {{Longo}}, \citenamefont
  {{Loparco}}, \citenamefont {{Lovellette}}, \citenamefont {{Lubrano}},
  \citenamefont {{Madejski}}, \citenamefont {{Massaro}}, \citenamefont
  {{Mayer}}, \citenamefont {{Mazziotta}}, \citenamefont {{McEnery}},
  \citenamefont {{Mehault}}, \citenamefont {{Michelson}}, \citenamefont
  {{Mignani}}, \citenamefont {{Mitthumsiri}}, \citenamefont {{Mizuno}},
  \citenamefont {{Moiseev}}, \citenamefont {{Monzani}}, \citenamefont
  {{Morselli}}, \citenamefont {{Moskalenko}}, \citenamefont {{Murgia}},
  \citenamefont {{Nakamori}}, \citenamefont {{Nemmen}}, \citenamefont {{Nuss}},
  \citenamefont {{Ohno}}, \citenamefont {{Ohsugi}}, \citenamefont {{Omodei}},
  \citenamefont {{Orienti}}, \citenamefont {{Orlando}}, \citenamefont
  {{Ormes}}, \citenamefont {{Paneque}}, \citenamefont {{Perkins}},
  \citenamefont {{Pesce-Rollins}}, \citenamefont {{Piron}}, \citenamefont
  {{Pivato}}, \citenamefont {{Rain{\`o}}}, \citenamefont {{Rando}},
  \citenamefont {{Razzano}}, \citenamefont {{Razzaque}}, \citenamefont
  {{Reimer}}, \citenamefont {{Reimer}}, \citenamefont {{Ritz}}, \citenamefont
  {{Romoli}}, \citenamefont {{S{\'a}nchez-Conde}}, \citenamefont {{Schulz}},
  \citenamefont {{Sgr{\`o}}}, \citenamefont {{Simeon}}, \citenamefont
  {{Siskind}}, \citenamefont {{Smith}}, \citenamefont {{Spandre}},
  \citenamefont {{Spinelli}}, \citenamefont {{Stecker}}, \citenamefont
  {{Strong}}, \citenamefont {{Suson}}, \citenamefont {{Tajima}}, \citenamefont
  {{Takahashi}}, \citenamefont {{Takahashi}}, \citenamefont {{Tanaka}},
  \citenamefont {{Thayer}}, \citenamefont {{Thayer}}, \citenamefont
  {{Thompson}}, \citenamefont {{Thorsett}}, \citenamefont {{Tibaldo}},
  \citenamefont {{Tibolla}}, \citenamefont {{Tinivella}}, \citenamefont
  {{Troja}}, \citenamefont {{Uchiyama}}, \citenamefont {{Usher}}, \citenamefont
  {{Vandenbroucke}}, \citenamefont {{Vasileiou}}, \citenamefont {{Vianello}},
  \citenamefont {{Vitale}}, \citenamefont {{Waite}}, \citenamefont {{Werner}},
  \citenamefont {{Winer}}, \citenamefont {{Wood}}, \citenamefont {{Wood}},
  \citenamefont {{Yamazaki}}, \citenamefont {{Yang}},\ and\ \citenamefont
  {{Zimmer}}}]{fermi13pi0}%
  \BibitemOpen
  \bibfield  {author} {\bibinfo {author} {\bibfnamefont {M.}~\bibnamefont
  {{Ackermann}}} {\it et~al.},\ }\bibfield  {title} {\enquote {\bibinfo {title}
  {{Detection of the Characteristic Pion-Decay Signature in Supernova
  Remnants}},}\ }\href {\doibase 10.1126/science.1231160} {\bibfield  {journal}
  {\bibinfo  {journal} {Science}\ }\textbf {\bibinfo {volume} {339}},\ \bibinfo
  {pages} {807} (\bibinfo {year} {2013})},\ \Eprint
  {http://arxiv.org/abs/1302.3307}{arXiv:1302.3307}\BibitemShut {NoStop}%
\bibitem [{\citenamefont {{Jogler}}\ and\ \citenamefont
  {{Funk}}(2016)}]{Jogler16w51c}%
  \BibitemOpen
  \bibfield  {author} {\bibinfo {author} {\bibfnamefont {T.}~\bibnamefont
  {{Jogler}}} and\ \bibinfo {author} {\bibfnamefont {S.}~\bibnamefont
  {{Funk}}},\ }\bibfield  {title} {\enquote {\bibinfo {title} {{Revealing W51C
  as a Cosmic Ray Source Using Fermi-LAT Data}},}\ }\href {\doibase
  10.3847/0004-637X/816/2/100} {\bibfield  {journal} {\bibinfo  {journal}
  {Astrophys. J.}\ }\textbf {\bibinfo {volume} {816}},\ \bibinfo {eid} {100}
  (\bibinfo {year} {2016})}\BibitemShut {NoStop}%
\bibitem [{\citenamefont {Raffelt}\ and\ \citenamefont
  {Stodolsky}(1988)}]{axionf}%
  \BibitemOpen
  \bibfield  {author} {\bibinfo {author} {\bibfnamefont {G.}~\bibnamefont
  {Raffelt}} and\ \bibinfo {author} {\bibfnamefont {L.}~\bibnamefont
  {Stodolsky}},\ }\bibfield  {title} {\enquote {\bibinfo {title} {Mixing of the
  photon with low-mass particles},}\ }\href {\doibase 10.1103/PhysRevD.37.1237}
  {\bibfield  {journal} {\bibinfo  {journal} {Phys. Rev. D}\ }\textbf {\bibinfo
  {volume} {37}},\ \bibinfo {pages} {1237} (\bibinfo {year}
  {1988})}\BibitemShut {NoStop}%
\bibitem [{\citenamefont {{Mirizzi}}\ and\ \citenamefont
  {{Montanino}}(2009)}]{axionf1}%
  \BibitemOpen
  \bibfield  {author} {\bibinfo {author} {\bibfnamefont {A.}~\bibnamefont
  {{Mirizzi}}} and\ \bibinfo {author} {\bibfnamefont {D.}~\bibnamefont
  {{Montanino}}},\ }\bibfield  {title} {\enquote {\bibinfo {title} {{Stochastic
  conversions of TeV photons into axion-like particles in extragalactic
  magnetic fields}},}\ }\href {\doibase 10.1088/1475-7516/2009/12/004}
  {\bibfield  {journal} {\bibinfo  {journal} {\jcap}\ }\textbf {\bibinfo
  {volume} {12}},\ \bibinfo {eid} {004} (\bibinfo {year} {2009})},\ \Eprint
  {http://arxiv.org/abs/0911.0015}{arXiv:0911.0015}\BibitemShut {NoStop}%
\bibitem [{\citenamefont {Jansson}\ and\ \citenamefont
  {Farrar}(2012)}]{Bfield1}%
  \BibitemOpen
  \bibfield  {author} {\bibinfo {author} {\bibfnamefont {R.}~\bibnamefont
  {Jansson}} and\ \bibinfo {author} {\bibfnamefont {G.~R.}\ \bibnamefont
  {Farrar}},\ }\bibfield  {title} {\enquote {\bibinfo {title} {A new model of
  the galactic magnetic field},}\ }\href
  {http://stacks.iop.org/0004-637X/757/i=1/a=14} {\bibfield  {journal}
  {\bibinfo  {journal} {The Astrophysical Journal}\ }\textbf {\bibinfo {volume}
  {757}},\ \bibinfo {pages} {14} (\bibinfo {year} {2012})}\BibitemShut
  {NoStop}%
\bibitem [{\citenamefont {{Sun}}\ \emph {{\it et~al.}}(2008)\citenamefont
  {{Sun}}, \citenamefont {{Reich}}, \citenamefont {{Waelkens}},\ and\
  \citenamefont {{En{\ss}lin}}}]{Bfield2}%
  \BibitemOpen
  \bibfield  {author} {\bibinfo {author} {\bibfnamefont {X.~H.}\ \bibnamefont
  {{Sun}}}, \bibinfo {author} {\bibfnamefont {W.}~\bibnamefont {{Reich}}},
  \bibinfo {author} {\bibfnamefont {A.}~\bibnamefont {{Waelkens}}}, and\
  \bibinfo {author} {\bibfnamefont {T.~A.}\ \bibnamefont {{En{\ss}lin}}},\
  }\bibfield  {title} {\enquote {\bibinfo {title} {{Radio observational
  constraints on Galactic 3D-emission models}},}\ }\href {\doibase
  10.1051/0004-6361:20078671} {\bibfield  {journal} {\bibinfo  {journal}
  {\aap}\ }\textbf {\bibinfo {volume} {477}},\ \bibinfo {pages} {573} (\bibinfo
  {year} {2008})},\ \Eprint
  {http://arxiv.org/abs/0711.1572}{arXiv:0711.1572}\BibitemShut {NoStop}%
\bibitem [{\citenamefont {Pshirkov}\ \emph {{\it et~al.}}(2011)\citenamefont
  {Pshirkov}, \citenamefont {Tinyakov}, \citenamefont {Kronberg},\ and\
  \citenamefont {Newton-McGee}}]{Bfield3}%
  \BibitemOpen
  \bibfield  {author} {\bibinfo {author} {\bibfnamefont {M.~S.}\ \bibnamefont
  {Pshirkov}}, \bibinfo {author} {\bibfnamefont {P.~G.}\ \bibnamefont
  {Tinyakov}}, \bibinfo {author} {\bibfnamefont {P.~P.}\ \bibnamefont
  {Kronberg}}, and\ \bibinfo {author} {\bibfnamefont {K.~J.}\ \bibnamefont
  {Newton-McGee}},\ }\bibfield  {title} {\enquote {\bibinfo {title} {Deriving
  the global structure of the galactic magnetic field from faraday rotation
  measures of extragalactic sources},}\ }\href
  {http://stacks.iop.org/0004-637X/738/i=2/a=192} {\bibfield  {journal}
  {\bibinfo  {journal} {The Astrophysical Journal}\ }\textbf {\bibinfo {volume}
  {738}},\ \bibinfo {pages} {192} (\bibinfo {year} {2011})}\BibitemShut
  {NoStop}%
\bibitem [{\citenamefont {{Atwood}}\ \emph {{\it et~al.}}(2013)\citenamefont
  {{Atwood}}, \citenamefont {{Albert}}, \citenamefont {{Baldini}},
  \citenamefont {{Tinivella}}, \citenamefont {{Bregeon}}, \citenamefont
  {{Pesce-Rollins}}, \citenamefont {{Sgr{\`o}}}, \citenamefont {{Bruel}},
  \citenamefont {{Charles}}, \citenamefont {{Drlica-Wagner}}, \citenamefont
  {{Franckowiak}}, \citenamefont {{Jogler}}, \citenamefont {{Rochester}},
  \citenamefont {{Usher}}, \citenamefont {{Wood}}, \citenamefont
  {{Cohen-Tanugi}},\ and\ \citenamefont {{S.~Zimmer for the Fermi-LAT
  Collaboration}}}]{pass8econf}%
  \BibitemOpen
  \bibfield  {author} {\bibinfo {author} {\bibfnamefont {W.}~\bibnamefont
  {{Atwood}}} {\it et~al.},\ }\bibfield  {title} {\enquote {\bibinfo {title}
  {{Pass 8: Toward the Full Realization of the Fermi-LAT Scientific
  Potential}},}\ }\href@noop {} {\bibfield  {journal} {\bibinfo  {journal}
  {ArXiv e-prints}\ } (\bibinfo {year} {2013})},\ \Eprint
  {http://arxiv.org/abs/1303.3514}{arXiv:1303.3514}\BibitemShut {NoStop}%
\bibitem [{\citenamefont {{Massaro}}\ \emph {{\it et~al.}}(2004)\citenamefont
  {{Massaro}}, \citenamefont {{Perri}}, \citenamefont {{Giommi}},\ and\
  \citenamefont {{Nesci}}}]{logp}%
  \BibitemOpen
  \bibfield  {author} {\bibinfo {author} {\bibfnamefont {E.}~\bibnamefont
  {{Massaro}}}, \bibinfo {author} {\bibfnamefont {M.}~\bibnamefont {{Perri}}},
  \bibinfo {author} {\bibfnamefont {P.}~\bibnamefont {{Giommi}}}, and\ \bibinfo
  {author} {\bibfnamefont {R.}~\bibnamefont {{Nesci}}},\ }\bibfield  {title}
  {\enquote {\bibinfo {title} {{Log-parabolic spectra and particle acceleration
  in the BL Lac object Mkn 421: Spectral analysis of the complete BeppoSAX wide
  band X-ray data set}},}\ }\href {\doibase 10.1051/0004-6361:20031558}
  {\bibfield  {journal} {\bibinfo  {journal} {\aap}\ }\textbf {\bibinfo
  {volume} {413}},\ \bibinfo {pages} {489} (\bibinfo {year} {2004})},\ \Eprint
  {http://arxiv.org/abs/astro-ph/0312260}{astro-ph/0312260}\BibitemShut
  {NoStop}%
\bibitem [{\citenamefont {{Chang}}\ \emph {{\it et~al.}}(2017)\citenamefont
  {{Chang}}, \citenamefont {{Ambrosi}}, \citenamefont {{An}}, \citenamefont
  {{Asfandiyarov}}, \citenamefont {{Azzarello}}, \citenamefont {{Bernardini}},
  \citenamefont {{Bertucci}}, \citenamefont {{Cai}}, \citenamefont
  {{Caragiulo}}, \citenamefont {{Chen}}, \citenamefont {{Chen}}, \citenamefont
  {{Chen}}, \citenamefont {{Chen}}, \citenamefont {{Cui}}, \citenamefont
  {{Cui}}, \citenamefont {{D'Amone}}, \citenamefont {{De Benedittis}},
  \citenamefont {{De Mitri}}, \citenamefont {{Di Santo}}, \citenamefont
  {{Dong}}, \citenamefont {{Dong}}, \citenamefont {{Dong}}, \citenamefont
  {{Dong}}, \citenamefont {{Donvito}}, \citenamefont {{Droz}}, \citenamefont
  {{Duan}}, \citenamefont {{Duan}}, \citenamefont {{Duranti}}, \citenamefont
  {{D'Urso}}, \citenamefont {{Fan}}, \citenamefont {{Fan}}, \citenamefont
  {{Fang}}, \citenamefont {{Feng}}, \citenamefont {{Feng}}, \citenamefont
  {{Fusco}}, \citenamefont {{Gallo}}, \citenamefont {{Gan}}, \citenamefont
  {{Gan}}, \citenamefont {{Gao}}, \citenamefont {{Gao}}, \citenamefont
  {{Gargano}}, \citenamefont {{Gong}}, \citenamefont {{Gong}}, \citenamefont
  {{Guo}}, \citenamefont {{Hu}}, \citenamefont {{Huang}}, \citenamefont
  {{Huang}}, \citenamefont {{Ionica}}, \citenamefont {{Jiang}}, \citenamefont
  {{Jiang}}, \citenamefont {{Jin}}, \citenamefont {{Kong}}, \citenamefont
  {{Lei}}, \citenamefont {{Li}}, \citenamefont {{Li}}, \citenamefont {{Li}},
  \citenamefont {{Li}}, \citenamefont {{Liang}}, \citenamefont {{Liang}},
  \citenamefont {{Liao}}, \citenamefont {{Liu}}, \citenamefont {{Liu}},
  \citenamefont {{Liu}}, \citenamefont {{Liu}}, \citenamefont {{Liu}},
  \citenamefont {{Liu}}, \citenamefont {{Liu}}, \citenamefont {{Loparco}},
  \citenamefont {{L{\"u}}}, \citenamefont {{Ma}}, \citenamefont {{Ma}},
  \citenamefont {{Ma}}, \citenamefont {{Ma}}, \citenamefont {{Ma}},
  \citenamefont {{Ma}}, \citenamefont {{Marsella}}, \citenamefont
  {{Mazziotta}}, \citenamefont {{Mo}}, \citenamefont {{Miao}}, \citenamefont
  {{Niu}}, \citenamefont {{Pohl}}, \citenamefont {{Peng}}, \citenamefont
  {{Peng}}, \citenamefont {{Qiao}}, \citenamefont {{Rao}}, \citenamefont
  {{Salinas}}, \citenamefont {{Shang}}, \citenamefont {{Shen}}, \citenamefont
  {{Shen}}, \citenamefont {{Shen}}, \citenamefont {{Song}}, \citenamefont
  {{Su}}, \citenamefont {{Su}}, \citenamefont {{Sun}}, \citenamefont {{Surdo}},
  \citenamefont {{Teng}}, \citenamefont {{Tian}}, \citenamefont {{Tykhonov}},
  \citenamefont {{Vagelli}}, \citenamefont {{Vitillo}}, \citenamefont {{Wang}},
  \citenamefont {{Wang}}, \citenamefont {{Wang}}, \citenamefont {{Wang}},
  \citenamefont {{Wang}}, \citenamefont {{Wang}}, \citenamefont {{Wang}},
  \citenamefont {{Wang}}, \citenamefont {{Wang}}, \citenamefont {{Wang}},
  \citenamefont {{Wang}}, \citenamefont {{Wang}}, \citenamefont {{Wang}},
  \citenamefont {{Wen}}, \citenamefont {{Wang}}, \citenamefont {{Wei}},
  \citenamefont {{Wei}}, \citenamefont {{Wei}}, \citenamefont {{Wu}},
  \citenamefont {{Wu}}, \citenamefont {{Wu}}, \citenamefont {{Wu}},
  \citenamefont {{Xi}}, \citenamefont {{Xia}}, \citenamefont {{Xin}},
  \citenamefont {{Xu}}, \citenamefont {{Xu}}, \citenamefont {{Xu}},
  \citenamefont {{Xue}}, \citenamefont {{Yang}}, \citenamefont {{Yang}},
  \citenamefont {{Yang}}, \citenamefont {{Yang}}, \citenamefont {{Yang}},
  \citenamefont {{Yao}}, \citenamefont {{Yu}}, \citenamefont {{Yuan}},
  \citenamefont {{Yue}}, \citenamefont {{Zang}}, \citenamefont {{Zhang}},
  \citenamefont {{Zhang}}, \citenamefont {{Zhang}}, \citenamefont {{Zhang}},
  \citenamefont {{Zhang}}, \citenamefont {{Zhang}}, \citenamefont {{Zhang}},
  \citenamefont {{Zhang}}, \citenamefont {{Zhang}}, \citenamefont {{Zhang}},
  \citenamefont {{Zhang}}, \citenamefont {{Zhang}}, \citenamefont {{Zhang}},
  \citenamefont {{Zhang}}, \citenamefont {{Zhang}}, \citenamefont {{Zhang}},
  \citenamefont {{Zhang}}, \citenamefont {{Zhao}}, \citenamefont {{Zhao}},
  \citenamefont {{Zhao}}, \citenamefont {{Zhou}}, \citenamefont {{Zhou}},
  \citenamefont {{Zhu}}, \citenamefont {{Zhu}},\ and\ \citenamefont
  {{Zimmer}}}]{DAMPE1}%
  \BibitemOpen
  \bibfield  {author} {\bibinfo {author} {\bibfnamefont {J.}~\bibnamefont
  {{Chang}}} {\it et~al.},\ }\bibfield  {title} {\enquote {\bibinfo {title}
  {{The DArk Matter Particle Explorer mission}},}\ }\href {\doibase
  10.1016/j.astropartphys.2017.08.005} {\bibfield  {journal} {\bibinfo
  {journal} {Astroparticle Physics}\ }\textbf {\bibinfo {volume} {95}},\
  \bibinfo {pages} {6} (\bibinfo {year} {2017})},\ \Eprint
  {http://arxiv.org/abs/1706.08453}{arXiv:1706.08453}\BibitemShut {NoStop}%
\bibitem [{\citenamefont {{DAMPE Collaboration}}(2017)}]{DAMPE2}%
  \BibitemOpen
  \bibfield  {author} {\bibinfo {author} {\bibnamefont {{DAMPE
  Collaboration}}},\ }\bibfield  {title} {\enquote {\bibinfo {title} {{Direct
  detection of a break in the teraelectronvolt cosmic-ray spectrum of electrons
  and positrons}},}\ }\href {\doibase 10.1038/nature24475} {\bibfield
  {journal} {\bibinfo  {journal} {Nature}\ }\textbf {\bibinfo {volume} {552}},\
  \bibinfo {pages} {63} (\bibinfo {year} {2017})},\ \Eprint
  {http://arxiv.org/abs/1711.10981}{arXiv:1711.10981}\BibitemShut {NoStop}%
\end{thebibliography}%

\clearpage

\widetext
\appendix
\section{Results for W44, W51C and Vela}
\label{sec7}
\begin{figure*}[b]
\centering
\includegraphics[width=0.45\textwidth]{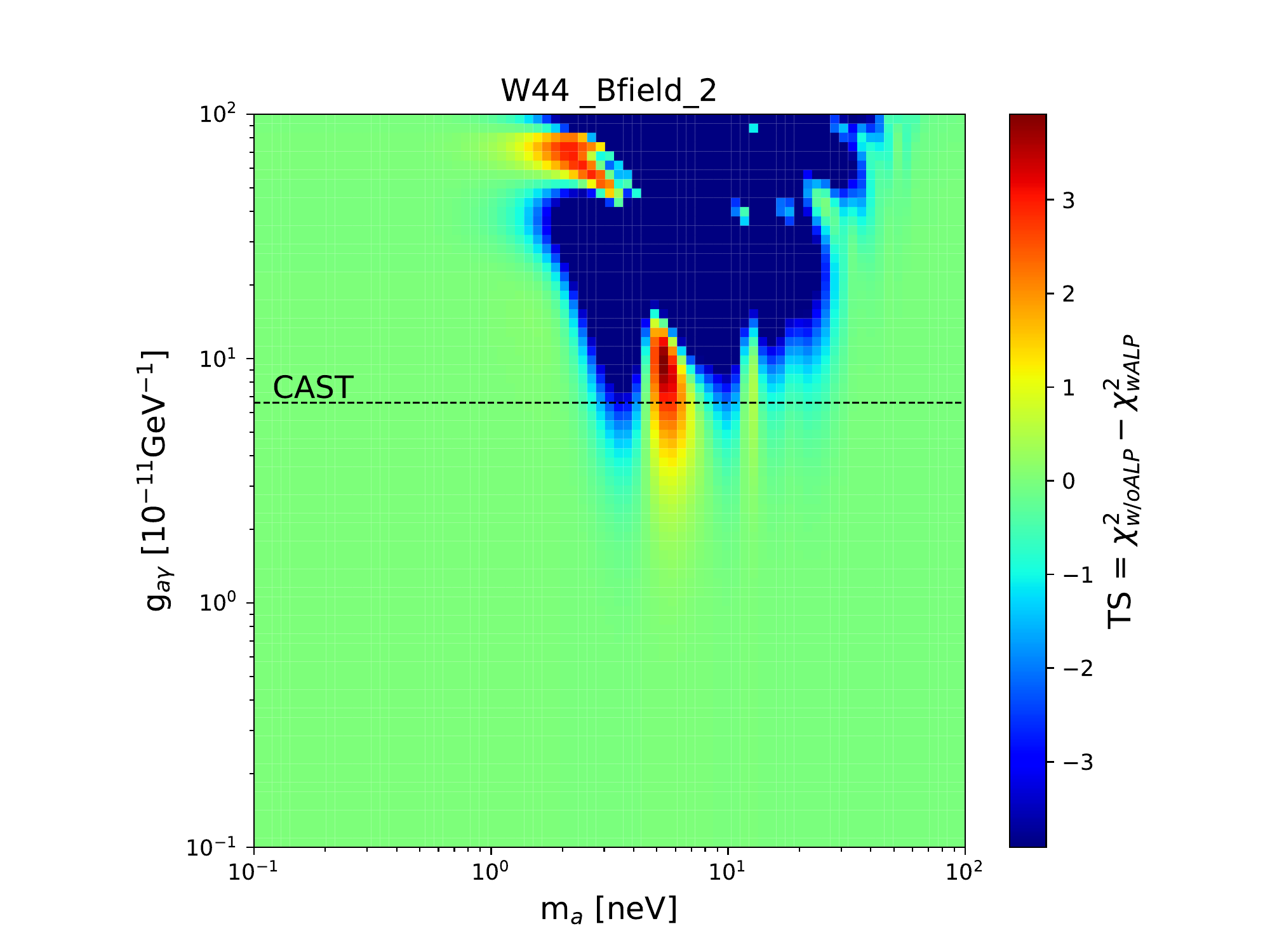}
\includegraphics[width=0.45\textwidth]{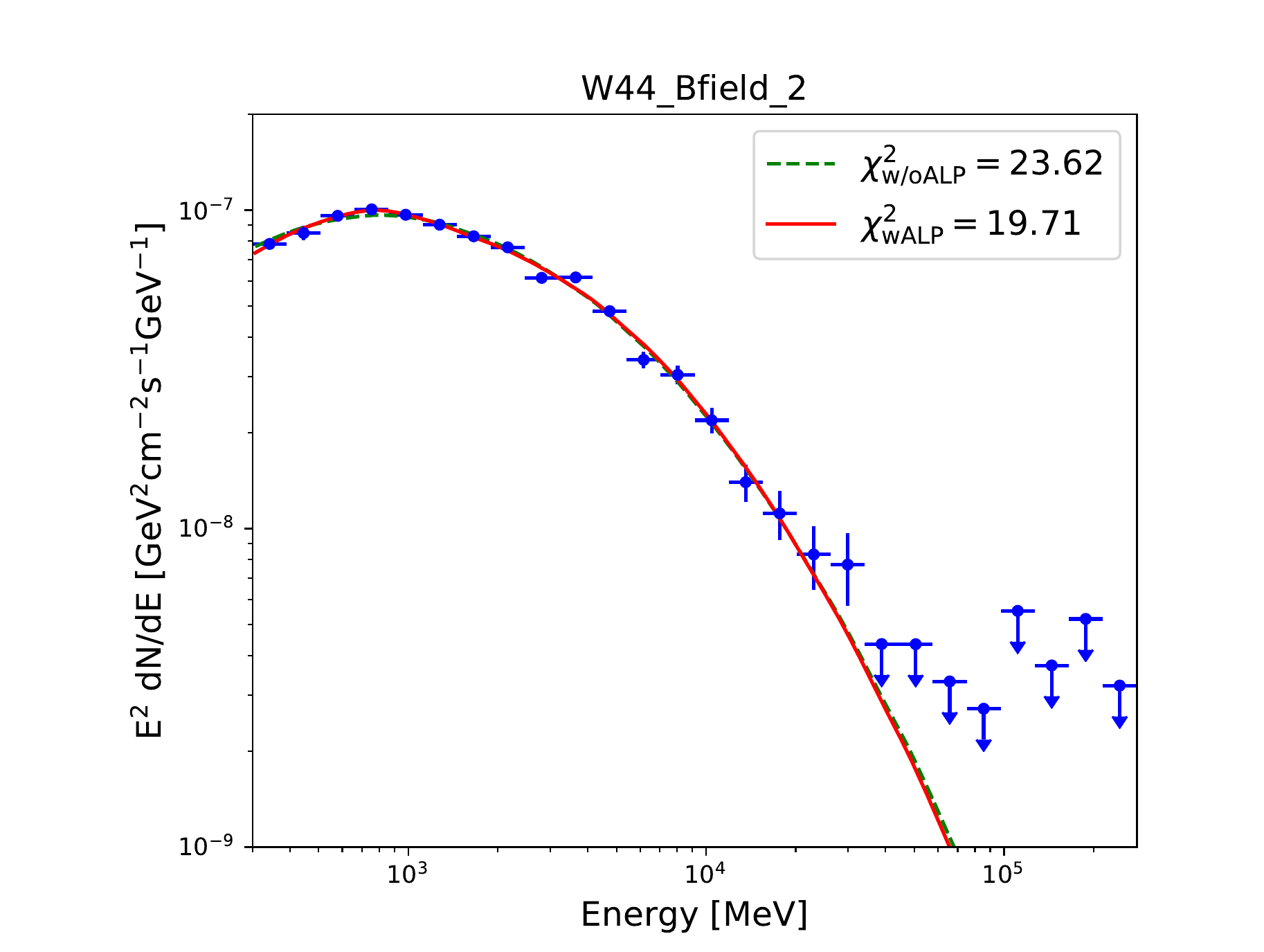}

\includegraphics[width=0.45\textwidth]{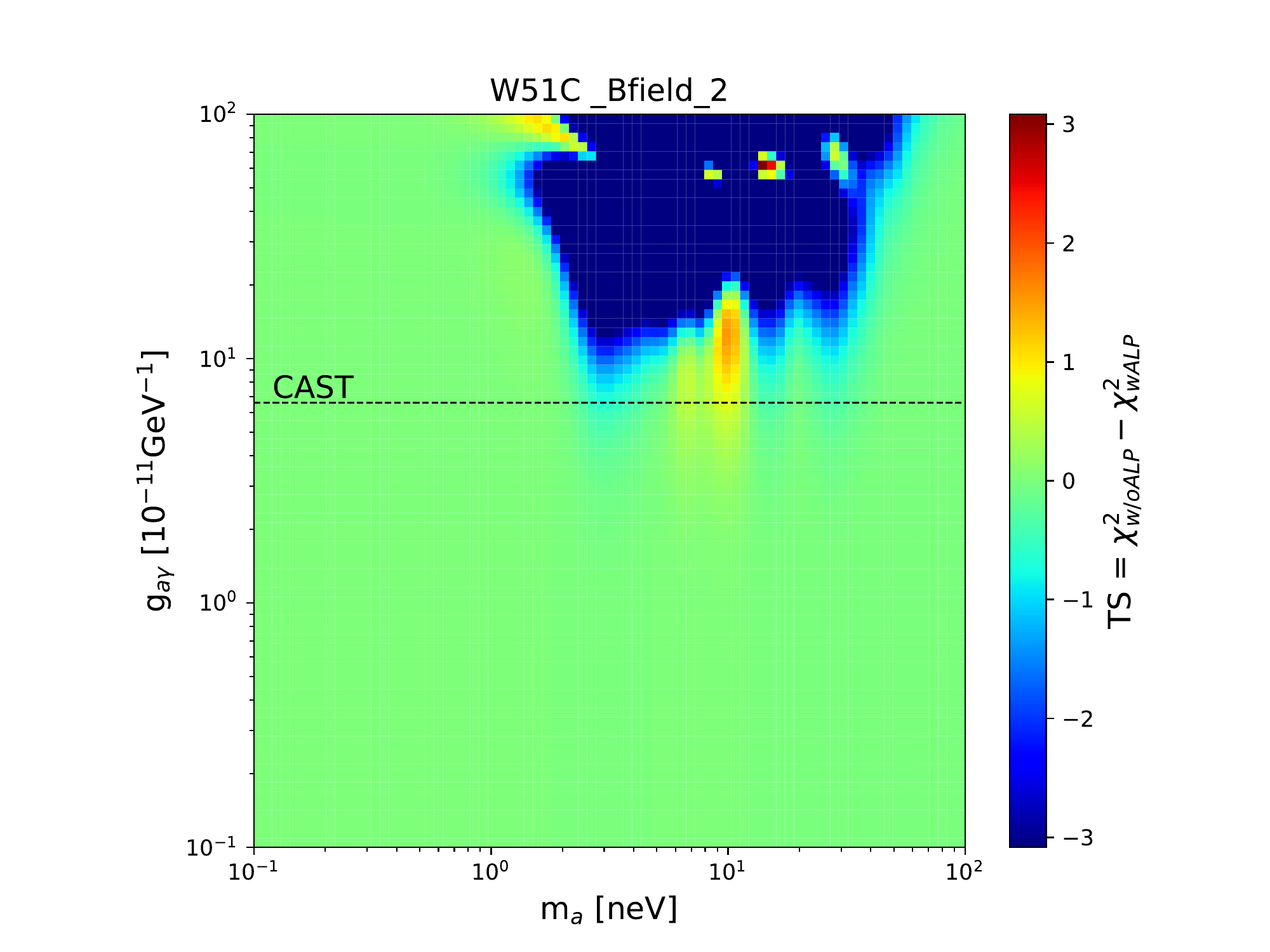}
\includegraphics[width=0.45\textwidth]{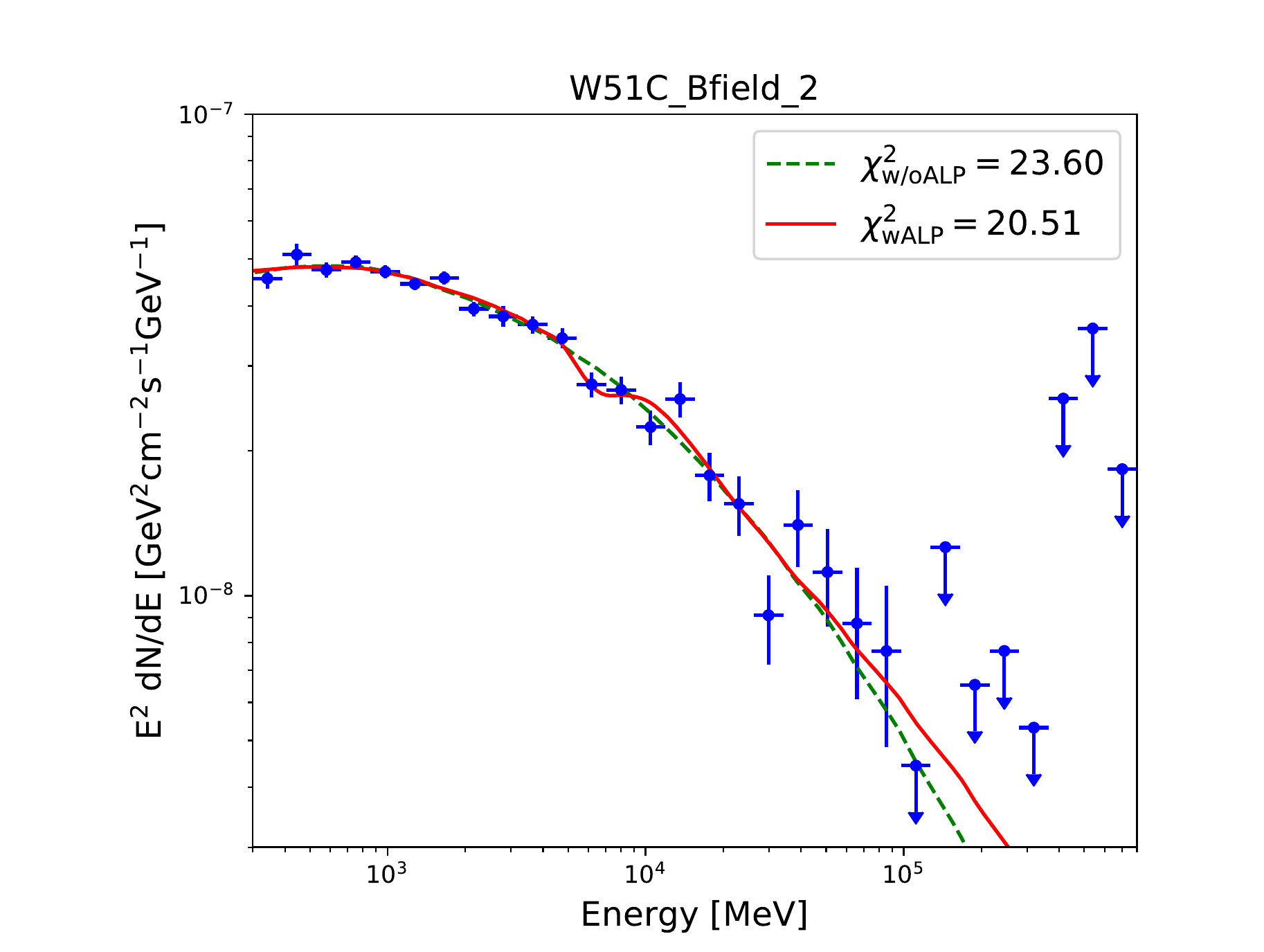}

\includegraphics[width=0.45\textwidth]{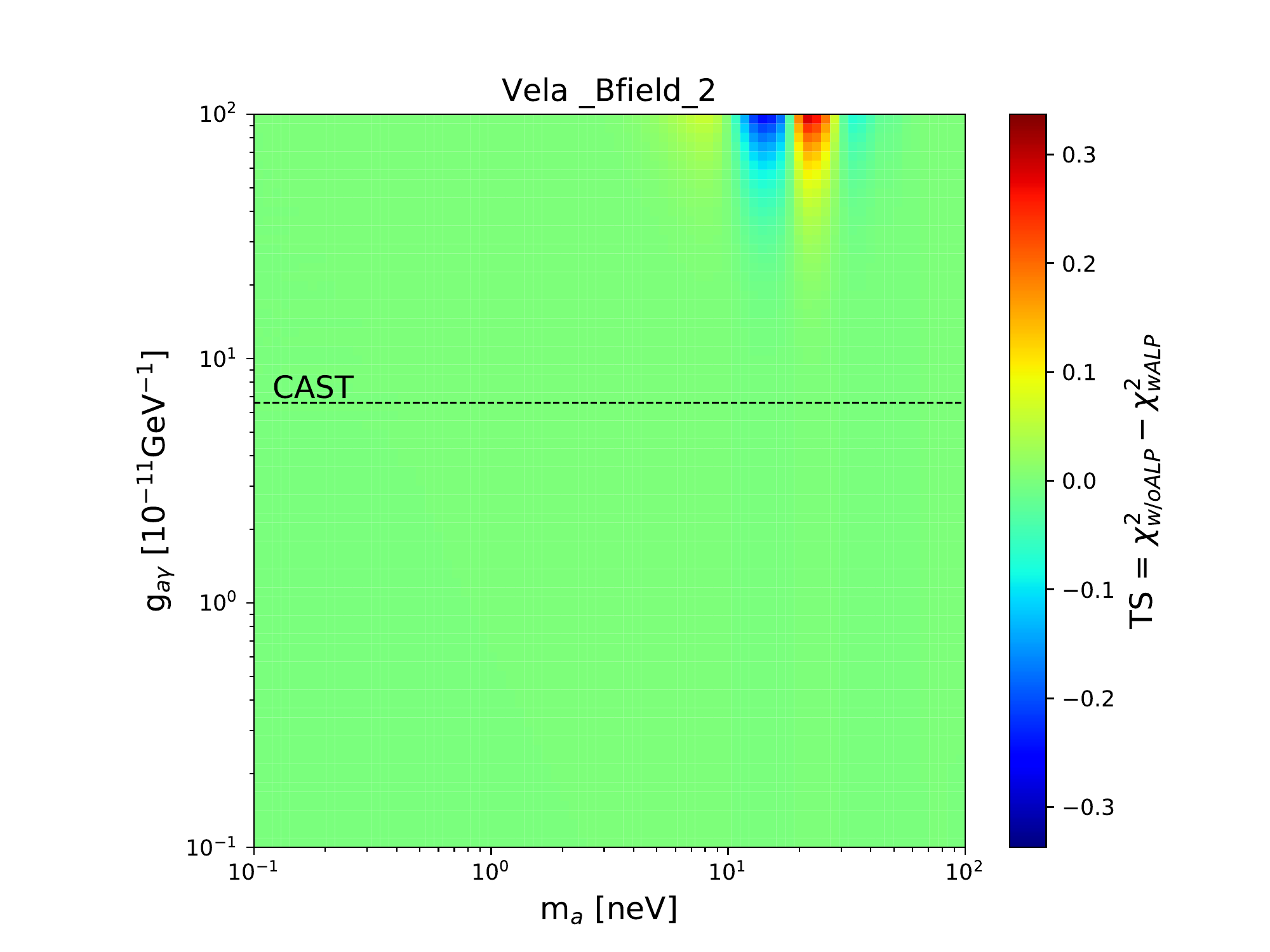}
\includegraphics[width=0.45\textwidth]{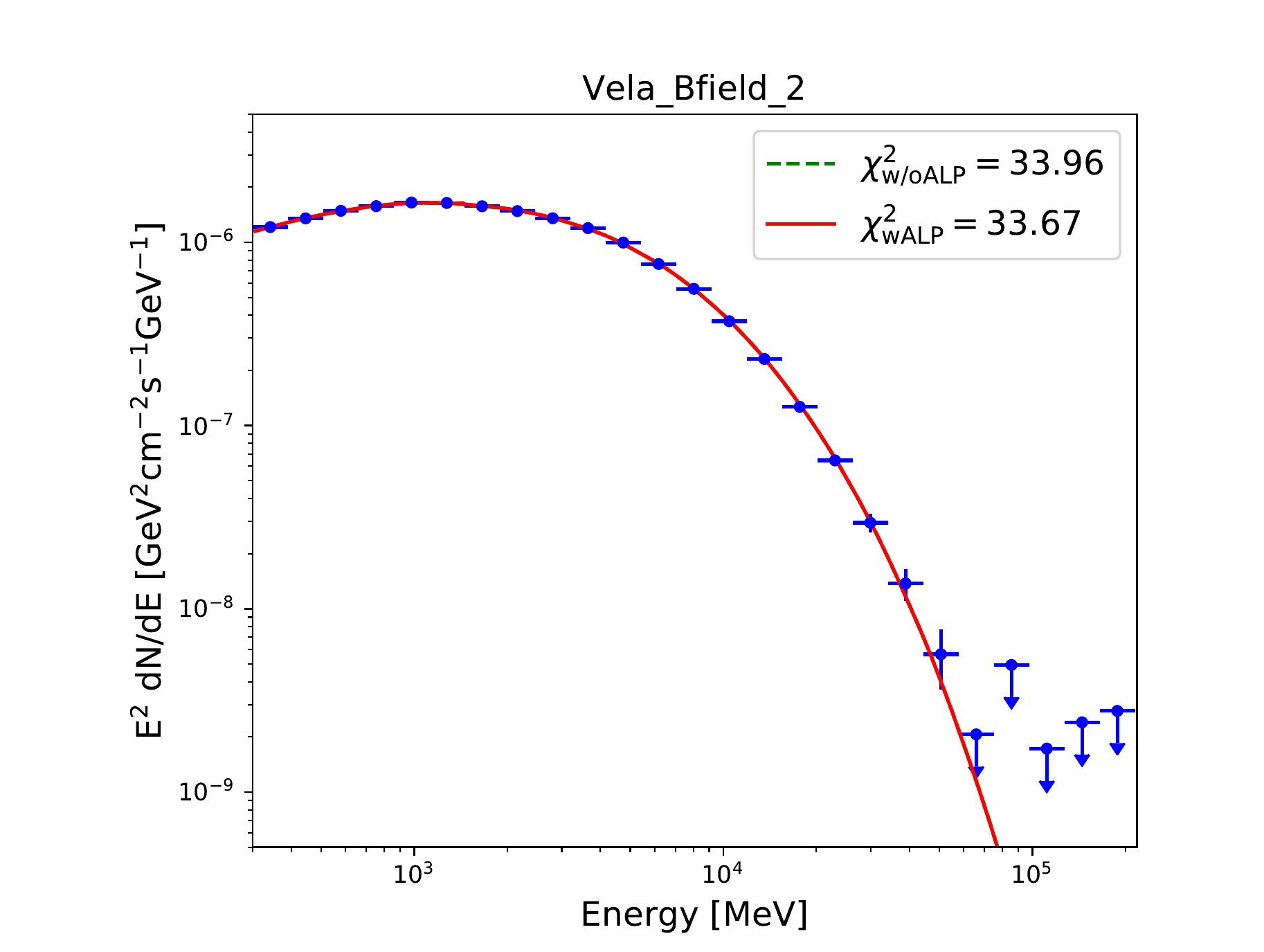}
\caption{Left column: The TS value as a function of ALP mass $m_{a}$ and photon-ALP coupling constant $g_{a\gamma}$. Right column: The Fermi-LAT SED and the best-fit spectra without and with photon-ALP oscillations. Upper, middle and lower panel are for W44, W51C and Vela, respectively. Bfield2 is adopted for all the results here.}
\label{fig:Other Results}
\end{figure*}

\clearpage

\end{document}